\pdfoutput=1 \pdfsuppresswarningpagegroup=1 
\documentclass[
 aps,%
 prb,%
 reprint, twocolumn,%
 groupedaddress,% 
 superscriptaddress,%
 amsfonts,%
 letterpaper,%
 footinbib,%
 noeprint%
]{revtex4-1}

\RequirePackage{amsmath,amssymb,bm}
\usepackage{pifont}
\RequirePackage{graphicx}
\RequirePackage[usenames,dvipsnames]{xcolor}

\usepackage{hyperref}
\hypersetup{
    colorlinks=true,
    citecolor=blue,
    linkcolor=magenta,
    filecolor=cyan,      
    urlcolor=magenta,
    pdfauthor = {Chuan Chen},
    pdfcreator = {\LaTeX\ and \flqq hyperref\frqq},
}

\usepackage{pdfpages}
\makeatletter\AtBeginDocument{\let\LS@rot\@undefined}\makeatother

\graphicspath{ {./Figures/} } 

\newcommand{\Fref}[1]{Fig.~\ref{#1}}
\newcommand{\Frefs}[1]{Figs.~\ref{#1}}
\newcommand{\Eqref}[1]{Eq.~\eqref{#1}}
\newcommand{\Eqrefs}[1]{Eqs.~\eqref{#1}}
\newcommand{\Tabref}[1]{Table~\ref{#1}}

\newcommand{\hc}{\text{H.\,c.}}

\newcommand{\cmark}{\ding{51}}

\newcommand{\tequiv}{{\,\equiv\,}}
\newcommand{\teq}{{\,=\,}}

\newcommand{\tapprox}{{\,\approx\,}}
\newcommand{\tright}{{\,\rightarrow\,}}
\newcommand{\tsimeq}{{\,\simeq\,}}
\newcommand{\tsim}{{\,\sim\,}}
\newcommand{\tlt}{{\,<\,}}
\newcommand{\tll}{{\,\ll\,}}
\newcommand{\tle}{{\,\le\,}}
\newcommand{\tgt}{{\,>\,}}
\newcommand{\tge}{{\,\ge\,}}
\newcommand{\tin}{{\,\in\,}}
\newcommand{\tminus}{{\,-\,}}
\newcommand{\tplus}{{\,+\,}}
\newcommand{\tpm}{{\,\pm\,}}

\newcommand{\tto}{{\,\to\,}}
\newcommand{\etal}{\emph{et al.}}

\begin{document}

\title{Correlated states of a triangular net of coupled quantum wires: 
Implications for the phase diagram of marginally twisted bilayer graphene}

\author{Chuan~Chen}
\affiliation{Centre for Advanced 2D Materials and Graphene Research Centre, 
National University of Singapore, Singapore 117546}
\affiliation{Max-Planck Institute for the Physics of Complex Systems,
01187 Dresden, Germany}

\author{A.~H.~{Castro Neto}}
\affiliation{Centre for Advanced 2D Materials and Graphene Research Centre, 
National University of Singapore, Singapore 117546}
\affiliation{Department of Physics, National University of Singapore, Singapore 
117542}

\author{Vitor~M.~Pereira}
\email[Correspondence to: ]{vpereira@nus.edu.sg}
\affiliation{Centre for Advanced 2D Materials and Graphene Research Centre, 
National University of Singapore, Singapore 117546}
\affiliation{Department of Physics, National University of Singapore, Singapore 
117542}

\date{\today}

\begin{abstract}
We explore in detail the electronic phases of a system consisting of three 
non-colinear arrays of coupled quantum wires, each rotated 120 degrees 
with respect to the next. A perturbative renormalization-group analysis reveals 
that multiple correlated states can be stabilized: a $s$-wave or $d 
\tpm id$ superconductor, a charge density wave insulator, a two-dimensional 
Fermi liquid, and a 2D Luttinger liquid (also known as smectic metal or 
sliding Luttinger liquid). The model provides an effective description of 
electronic interactions in small-angle twisted bilayer graphene and we discuss 
its implications in relation to the recent observation of correlated and 
superconducting ground states near commensurate densities in magic-angle 
twisted samples, as well as the ``strange metal'' behavior at finite 
temperatures as a natural outcome of the 2D Luttinger liquid phase.
\end{abstract}

\maketitle

% ------------------------------------------------------------------------------
% MAIN TEXT BEGINS
% ------------------------------------------------------------------------------

\section{Introduction}

The low-energy physics of interacting fermions in one dimension (1D) is 
determined by collective spin and charge density excitations that define what is 
known as Luttinger liquid (LL) behavior \cite{Haldane1981, Giamarchi2003, 
Fradkin2013}. Soon after high-temperature superconductivity was discovered in 
cuprate oxides \cite{Bednorz1986}, it was proposed that the charges added upon 
doping a Mott insulator could end up distributed in stripes \cite{Tranquada1994, 
Zaanen1989, Machida1989, Kato1990, Emery1999}. This led Anderson \etal{} to 
suggest that confined fermionic excitations in such presumed LL arrays (the 
stripes) could explain the non-Fermi-liquid nature of cuprates' ``normal'' state 
\cite{Strong1994}. 
Since then, theoretical investigation has assessed whether LL behavior can 
emerge in higher dimensions, especially in 2D \cite{Strong1994, Wen1990, 
Emery2000, Vishwanath2001, Mukhopadhyay2001, Mukhopadhyay2001a}, the natural 
route to that having been to study systems of coupled LLs in different guises. 
It is now known, for example, that, in an array of parallel LLs, marginal 
inter-wire density-density and current-current interactions lead to strong 
transverse charge-density fluctuations at incommensurate wave vectors which can 
frustrate electron crystallization and indeed stabilize a LL state, commonly 
designated ``smectic metal'' or ``sliding Luttinger liquid'' state 
\cite{Emery2000, Vishwanath2001, Mukhopadhyay2001, Mukhopadhyay2001a}.
However, previous work has been limited to exploring consequences of couplings 
among either one or two perpendicularly crossed arrays, without ever considering 
LLs interlinked in the form of a triangular net, possibly for lack of a 
realistic representative system.

The recent discovery of strongly correlated physics in marginally twisted 
bilayer graphene (MTBG) near the magic angle $\theta \tapprox 1.1^{\circ}$ 
\cite{Cao2018, Cao2018a} set off a flurry of interest in the origin of the 
observed insulating and superconducting (SC) phases. (For the purposes of this 
work, MTBG refers to bilayers twisted by $\tsim 1^{\circ}$ or less, including 
the first magic angle.)
It had been previously suggested that, at magic angles, the quasi-flatness of 
the electronic bands closest to the undoped Fermi level could promote electronic 
instabilities \cite{SuarezMorell2010, Li2010, Bistritzer2011, 
LopesdosSantos2012}. 
The development of effective tight-binding models for those bands 
\cite{LopesdosSantos2007, Koshino2018, Po2018, Kang2018} enabled, on the one 
hand, predictions of possible broken symmetries arising from weak-coupling 
mechanisms, such as Fermi surface nesting or enhanced density of states 
\cite{Liu2018, Isobe2018, You2019, Po2018, Laksono2018}; on the other hand, it 
revealed trilobed Wannier functions centered at the AB/BA positions 
\cite{Koshino2018, Po2018, Kang2018}, which has in turn motivated 
strong-coupling perspectives based on extended and non-conventional Hubbard-type 
interactions \cite{Xu2018, Xiao2018, Gu2019}. Electronic interactions are an 
undisputed factor given that the ratio of the local Coulomb integral to 
bandwidth is estimated in the range $U/w \tsim 5\text{--}10$ 
\cite{Koshino2018}. 
The extremely large Moir\'e unit cells involved ($\tsim 172$\,nm$^2$, about 100 
times those of canonical Mott-insulators like cuprates) has also prompted the
suggestion that the insulating phase can be a Wigner crystal, consistently 
with the extremely low densities, and the emergence of SC a result of its 
melting \cite{Padhi2018, Padhi2019}. 

Evidence accumulated from recent experiments and theoretical work motivates 
a different perspective over the effective electronic model governing 
correlations in MTBG, which we develop in this paper. 
We note, first, that, by allowing more than the minimum two orbitals 
\cite{Po2018} per Moir\'e unit cell, Carr \etal{} have recently shown that the 
weight of the Bloch states belonging to the flat-band sector is overwhelmingly 
distributed among Wannier functions situated at the AA positions \emph{and} at 
the AB/BA \emph{domain boundaries} \cite{Carr2019b}.
Second, it is well known that, in the presence of perpendicular electrical 
fields, AB-BA domain boundaries host protected helical modes \cite{Martin2008, 
Kindermann2012, Zhang2013, Vaezi2013, Koshino2013}. MTBG accommodates well 
defined, intrinsic, and periodically alternating AB/BA regions 
\cite{LopesdosSantos2012} whose network of boundaries was shown to likewise 
support the propagation of such confined states \cite{San-Jose2013, Efimkin2018, 
Pal2018a}. Moreover, since AB is favored against AA stacking, a considerable 
atomic relaxation within the Moir\'e unit cell maximizes the AB/BA regions, 
leaving sharply defined, atomic-scale domain boundaries \cite{Alden2013, 
Nam2017, Anelkovic2017, Gargiulo2018, Lucignano2019, Walet2019a}. Crucially, 
there is now unequivocal spectroscopic \cite{Alden2013, Ju2015, Yin2016, 
Huang2018, Sunku2018} and transport \cite{Ju2015, Rickhaus2018, Yoo2018, Xu2019} 
evidence of the reality of this network of 1D modes in MTBG, including in 
single-gated devices.

There are additional hints that warrant a description in terms of such 
``network of linked quantum wires'' to describe the observed correlated 
behavior: the confinement of electrons to 1D naturally boosts correlations; a 
phase diagram similar to that of magic-angle samples arises at other twist 
angles under pressure \cite{Yankowitz2019}, in line with the expectation that an 
inter-layer coupling enhanced by pressure would amplify the lattice relaxation, 
in turn defining sharp domain boundaries and the emergence of the wire network 
for more generic twists \cite{Rickhaus2018, Yoo2018, Xu2019}. 

All the above aspects and observations call for an investigation of the 
implications of a coupled-wire description of the low-energy physics of MTBG. Wu 
\etal{} have recently advanced arguments to justify the insulating and SC phases 
in such a scenario, but only considering coupling at the wire intersections and 
spin isotropic interactions within each wire \cite{Wu2019}. 
By generalizing to MTBG the approach developed to study sliding LL phases, we 
scrutinize not only the competition among SC and charge-density wave (CDW) 
states, but also the emergence of Fermi liquid (FL) and sliding LL phases that, 
stabilized by inter-wire interactions, might explain the experimental 
progression of MTBG from an insulator to SC to a metal as density deviates 
from 
commensurate fillings. The SC order parameter acquires either $s$ or $d\tpm id$ 
symmetry, depending on the Josephson coupling at the wire crossings. Moreover, 
it suggests that these correlated phases could also happen in ``marginally'' 
($\theta \tll 1^\circ$) twisted bilayer graphene.
Finally, we emphasize that, independently of its direct relevance to MTBG, 
this work reports the first detailed study of coupled quantum wires and sliding 
LL phases in a triangular net geometry.

% ------------------------------------------------------------------------------
% FIGURE
% ------------------------------------------------------------------------------
\begin{figure}
\centering
\includegraphics[width=0.47\textwidth]{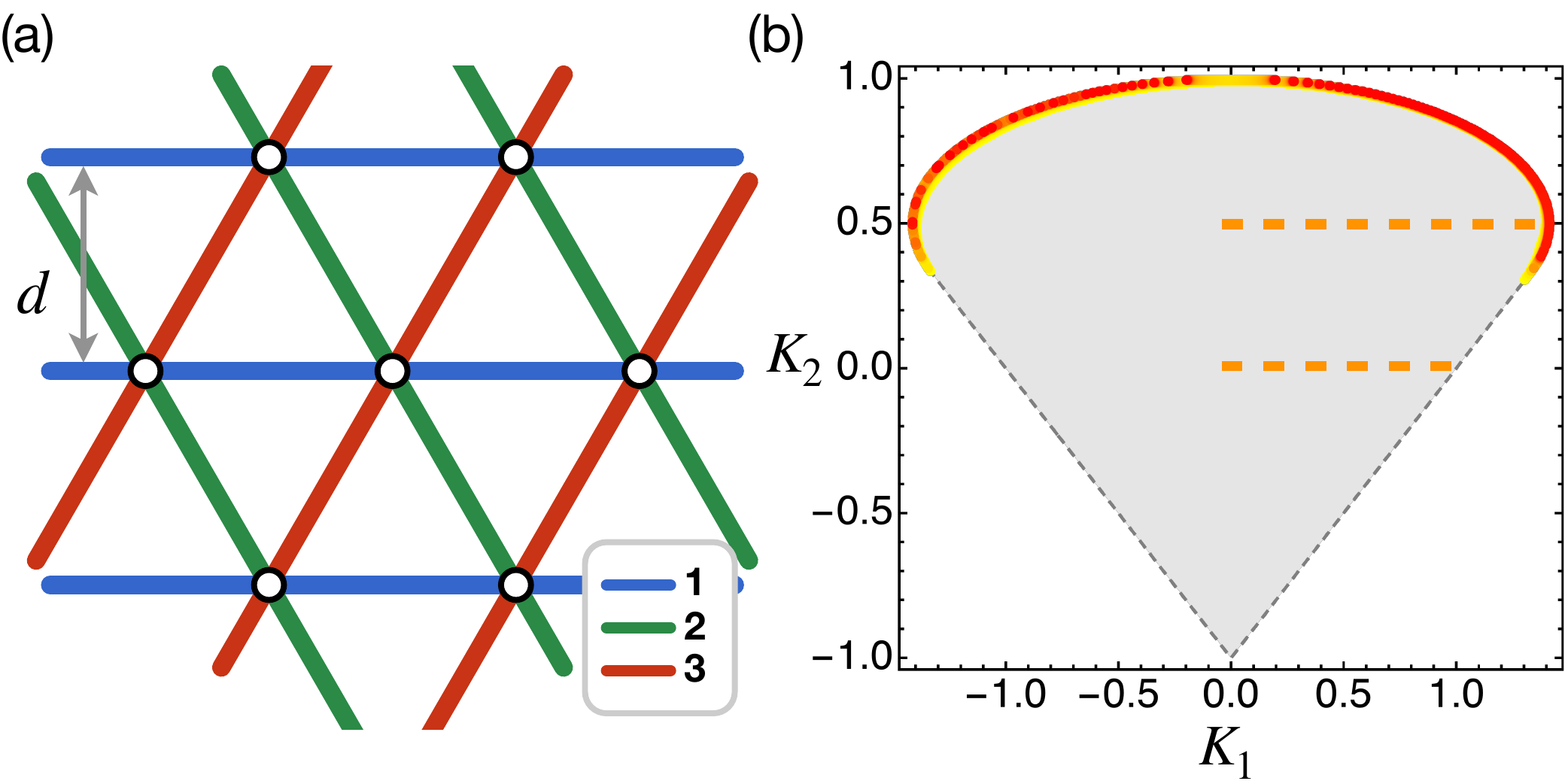}
\caption{(a) Schematic of the local net spanned by the three coupled arrays of 
quantum wires, labeled $\{1,2,3\}$. One \emph{array} consists of a set of 
parallel, identically colored wires. The superposition of the three arrays 
reproduces the net of AB-BA domain boundaries in MTBG.
(b) Domain of stability. The array's effective Luttinger parameter, 
$\kappa(k_\perp)$, is positive in the shaded domain for all $k \tin 
(-\pi/d,\pi/d)$. The LL phase exists \emph{only} in the red-to-yellow region 
surrounding the upper boundary. Dashed lines indicate the cuts chosen to 
generate the phase diagrams in \Fref{fig:phase-diagram}.
}
\label{fig:schematic}
\end{figure}
% ------------------------------------------------------------------------------

% ------------------------------------------------------------------------------
\section{Coupled-wire model}
% ------------------------------------------------------------------------------

\subsection{Interconnected Luttinger liquid array}

Mirroring the experimental network of AB-BA boundaries, we consider three 
families of quantum wires, each family consisting of an array of parallel 
wires depicted by the same color in \Fref{fig:schematic}(a). Although each link 
in MTBG supports two helical channels per spin along the forward and backward 
direction, we study the simplified case of a single mode per wire, which should 
capture the essential physics at play.
Each independent wire is a LL whose excitations are best described in the boson 
formalism by a separated spin ($s$) and charge ($c$) Hamiltonian 
(see Appendix for the convention of bosonization)
\cite{Haldane1981, Giamarchi2003, Fradkin2013},
\begin{align}\label{eq:LL}
  H = \sum_{\alpha = c,s} \int\!dx \ \frac{v_{\alpha}}{2}
  \left[ \frac{1}{K_{\alpha}} (\partial_x \theta_{\alpha})^2
  + K_{\alpha} (\partial_x \phi_{\alpha})^2 \right]\!,
\end{align}
where $\theta_\alpha(x)$ and $\phi_\alpha(x)$ are the conventional phase-field 
operators, $K_{c/s}$ is the (inverse) charge/spin Luttinger parameter 
($K_c\,\text{\small$\gtrless$}\,1$ for repulsive/attractive interactions), and 
$v_\alpha$ defines the velocity of each excitation. Spin back-scattering adds 
the term $2 g_s/(2\pi \alpha)^2\int\!dx \cos\left( 2\sqrt{2\pi} \phi_s 
\right)$, where $\alpha$ is the (length) cutoff of the theory, to Hamiltonian 
\eqref{eq:LL}, leading to a sine-Gordon-like action and spin couplings that 
flow according to the equations \citep{Giamarchi2003}
\begin{equation}\label{eq:sine-Gordon}
  \frac{dg_s}{dl} = 2(1-\frac{1}{K_s})g_s,
  \qquad
  \frac{dK_s}{dl} = \frac{g_s^2}{2\pi^2 v_s^2}.
\end{equation}
As a result, when there is a spin gap, $K_s \tright \infty$. 

To describe each periodic array of parallel wires separated by $d$ as in 
\Fref{fig:schematic}, one must include the long-wavelength (charge) 
density-density and current-current interactions among wires in the fixed-point 
Hamiltonian, as first noted by Emery \etal\ \cite{Emery2000}. It can then be 
shown 
that the charge part of the action reads, in Fourier space, 
\begin{multline}\label{eq:action-array}
  S = \sum_{q} \frac{v(k_{\perp}) k^2}{2} \left[ 
    \frac{|\theta_{c,q}|^2}{\kappa(k_{\perp})} + \kappa(k_{\perp}) 
    |\phi_{c,q}|^2 \right]
    \\
    + i \omega_n k \phi_{c,q}^* \theta_{c,q},
\end{multline}
where $q \tequiv (\omega_n,k,k_{\perp})$ and $k/k_{\perp}$ is the momentum 
along/perpendicular to the wires \cite{Emery2000, Vishwanath2001, 
Mukhopadhyay2001}. The spin part, for $\theta_s$ and $\phi_s$, is obtained by 
replacing $[ v(k_{\perp}), \kappa(k_{\perp}) ] \to [ v_s, K_s ]$.
Direct comparison with \Eqref{eq:LL} shows that an array of LLs effectively 
behaves as a LL, the net effect of the inter-wire coupling being a Luttinger 
parameter $\kappa$ that is now a $2\pi/d$-periodic function of $k_{\perp}$. 
This periodicity justifies a Fourier expansion, 
\begin{equation}
  \kappa(k_{\perp}) = K_0[ 
    1 + K_1 \cos(k_{\perp}d) + K_2 \cos(2k_{\perp}d) + \dots],
  \label{eq:kappa}
\end{equation}
which we shall use below with $K_{0,1,2}$ as free parameters 
\cite{Vishwanath2001, Mukhopadhyay2001, Mukhopadhyay2001a}.

Each of the three LL arrays depicted in \Fref{fig:schematic} is assigned a 
(superscript) label $j \tin \{1,2,3\}$. Within each array $j$, we consider 
the single-electron hopping ($t_\perp$) between nearest-neighboring wires, as 
well as inter-wire CDW ($\mathcal{V}_n$) and SC ($\mathcal{J}_n$) singlet 
interactions between $n$-th neighboring wires. These are described, 
respectively, by the \emph{intra}-array, \emph{inter}-wire couplings
\begin{subequations}\label{eq:intra-array}
\begin{align}
&\mathcal{H}^j_{h} = t_{\perp} \sum_{l,\sigma} \sum_{\nu = \pm1}
    \psi^{j \dagger}_{l,\nu,\sigma} \psi^j_{l+1,\nu,\sigma} + \hc,  \\
&\mathcal{H}^j_{c,n} = \mathcal{V}_n \sum_{l,\sigma,\sigma',\nu}
    \psi^{j \dagger}_{l,\nu,\sigma}\psi^j_{l,-\nu,\sigma} 
    \psi^{j \dagger}_{l+n,-\nu,\sigma'} \psi^j_{l+n,\nu,\sigma'}, \\
&\mathcal{H}^j_{sc,n} = \mathcal{J}_n \sum_{l,\mu,\nu}
    \psi^{j \dagger}_{l,\mu,\uparrow}\psi^{j \dagger}_{l,-\mu,\downarrow}
    \psi^j_{l+n,\nu,\downarrow}\psi^j_{l+n,-\nu,\uparrow} + \hc,
\end{align}
\end{subequations}
where $\psi^{i}_{l,\pm1,\sigma}$ is the field operator for a right/left-moving 
electron of spin $\sigma$ in the $l$-th wire of array $i$. The wires are 
also coupled at each intersection [white dots in \Fref{fig:schematic}(a)], 
requiring us to consider the additional \emph{inter}-array hopping, CDW and SC 
interactions:
\begin{subequations}\label{eq:inter-array}
\begin{align}
  &H^{i,j}_{h} = t \sum_{l,m,\sigma}\sum_{\mu,\nu}
    \psi^{i \dagger}_{l,\mu,\sigma} \psi^{j}_{m,\nu,\sigma} + \hc, \\
  &H^{i,j}_{c} = \mathcal{V}_0 \sum_{l,m} \sum_{\sigma,\sigma'} \sum_{\mu,\nu}
    \psi^{i \dagger}_{l,\mu,\sigma}\psi^{i}_{l,\mu,\sigma} 
    \psi^{j \dagger}_{m,\nu,\sigma'} \psi^{j}_{m,\nu,\sigma'}, \\
  &H^{i,j}_{sc} = \mathcal{J}_0 \! \sum_{l,m,\mu,\nu} \!
    \psi^{i \dagger}_{l,\mu,\uparrow}\psi^{i \dagger}_{l,-\mu,\downarrow}
    \psi^{j}_{m,\nu,\downarrow} \psi^{j}_{m,-\nu,\uparrow} + \hc.
\end{align}
\end{subequations}

% ------------------------------------------------------------------------------
\subsection{Renormalization group equations}
% ------------------------------------------------------------------------------

Once all the couplings  in \Eqrefs{eq:intra-array} and \eqref{eq:inter-array} 
are written in terms of the bosonic fields, we proceed by developing a 
perturbative renormalization group (RG) analysis. To the lowest order, the flow 
equations for the hopping ($t$, $t_\perp$), CDW ($\mathcal{V}_n$), and SC 
($\mathcal{J}_n$) coupling parameters read:
\begin{subequations} \label{eq:RG-flow}
\begin{align}
  &\frac{d\mathcal{V}_n}{dl} = (2-\delta_{n,0}-\frac{1}{K_s}-\Delta_{C,n}) 
    \mathcal{V}_n, \\
  &\frac{d\mathcal{J}_n}{dl} = (2-\delta_{n,0}-\frac{1}{K_s}-\Delta_{S,n}) 
    \mathcal{J}_n, \\
  &\frac{dt_{\perp}}{dl} = \left[ 2-\frac{1}{4} (K_s+\frac{1}{K_s})
    - \frac{1}{4}(\Delta_{C,1}+\Delta_{S,1}) \right] t_{\perp}, \\
  &\frac{dt}{dl} = \left[ 1-\frac{1}{4} (K_s+\frac{1}{K_s})
    - \frac{1}{4} ( \Delta_{C,0}+\Delta_{S,0} ) \right] t, 
    \label{eq:RG-flow-t}
\end{align}
\end{subequations}
where, 
\begin{subequations}\begin{align}
  \Delta_{C,n} & \equiv \int_{-\pi}^{\pi} \frac{dk}{2\pi} [ 1 - (1-\delta_{n,0}) 
  \cos(nk) ] \frac{1}{\kappa(k/d)} ,
  \\
  \Delta_{S,n} & \equiv \int_{-\pi}^{\pi} \frac{dk}{2\pi} [ 1 - (1-\delta_{n,0}) 
  \cos(nk) ] \, \kappa(k/d).
\end{align}\end{subequations}
It is physically reasonable to expect the intra-array couplings to decay 
rapidly so, henceforth, we only consider intra-array CDW and SC interactions up 
to second-neighbors. As for $\kappa(k_{\perp})$, in line with Vishwanath and 
Carpentier \cite{Vishwanath2001}, we truncate its Fourier expansion at the 
second order. Furthermore, in order to have a stable theory, $\kappa(k_{\perp})$ 
must be positive for $k_{\perp} \in (-\pi/d, \pi/d)$, which constrains $K_0 \tgt 
0$ and $(K_1, K_2)$ to the shaded domain shown in \Fref{fig:schematic}(b).

% ------------------------------------------------------------------------------
% TABLE
% ------------------------------------------------------------------------------
\begin{table}
\centering
\begin{tabular}{| c || c | c | c | c | c |} 
\hline
Coupling & $( 0, K_{-} ) $ & $( K_{-}, 1/2 )$ & $( 1/2, 1 )$ &
$( 1, K_{+} )$ & $( K_{+}, \infty )$ \\ [0.5ex] 
\hline
\multicolumn{6}{c}{intra-array (among parallel wires)} \\
\hline
$\mathcal{V}_n$ (CDW) & -- & -- & \cmark & \cmark & \cmark \\ 
$\mathcal{J}_n$ (SC) & -- & -- & \cmark & \cmark & \cmark \\ 
$t_\perp$ (hop) & -- & \cmark & \cmark & \cmark & -- \\[0.5ex]
\hline
\multicolumn{6}{c}{inter-array (at wire crossings)} \\
\hline
$\mathcal{V}_0$ (CDW) & -- & -- & -- & \cmark & \cmark \\ 
$\mathcal{J}_0$ (SC) & -- & -- & -- & \cmark & \cmark \\ 
$t$ (hop) & -- & -- & -- & -- & -- \\[0.5ex] 
\hline
\end{tabular}
\caption{
Relevance of each coupling for the different ranges of the spin Luttinger 
parameter ($K_s$) specified in the first row. The symbol \cmark{} means that a 
coupling \emph{may} be relevant while ``--'' indicates it is \emph{always} 
irrelevant within that interval of $K_s$. $K_{-} \tequiv 3 \tminus 2\sqrt{2} 
\tsimeq 0.17$, $K_{+} \tequiv 3 \tplus 2\sqrt{2} \tsimeq 5.83$. 
}
\label{table:phases}
\end{table}
% ------------------------------------------------------------------------------

At this level of approximation, the RG equations \eqref{eq:RG-flow} are 
independent. The relevancy of the different couplings can thus be immediately 
established and is summarized in \Tabref{table:phases}. 
Since $\kappa$ is strictly positive, $\Delta_{C,0}+\Delta_{S,0} \tge 2$ which, 
according to \Eqref{eq:RG-flow-t}, implies that the single-electron hopping at 
the wire intersections ($t$) is, at most, marginal if $K_s \teq 1$ and 
$\kappa(k/d) \teq 1$ for all $k$; it is otherwise irrelevant in nearly the whole 
phase space. This justifies considering $t$ globally irrelevant and, 
accordingly, it will be ignored in the subsequent analysis. 
Similarly, one can see that $\Delta_{C,1}+\Delta_{S,1} \tge 2$ so that the 
intra-array hopping ($t_{\perp}$) may be relevant when $3-2\sqrt{2} \tle K_s 
\tle 3+2\sqrt{2}$. The CDW and SC couplings are relevant only if $K_s \tgt 1/2$ 
in the intra-array case ($\mathcal{V}_{1,2}$ and $\mathcal{J}_{1,2}$), while the 
corresponding inter-array couplings ($\mathcal{V}_0$ and $\mathcal{J}_0$) are 
relevant for $K_s \tgt 1$.

Up to this point, the spin Luttinger parameter $K_s$ has been considered free; 
\Tabref{table:phases} thus covers the most general scenario in relation to the 
possible magnetic phases. However, addition of the spin backscattering term 
mentioned earlier to \Eqref{eq:LL} makes $K_s$ a running coupling, governed by 
the flow \Eqrefs{eq:sine-Gordon}. The solution where $K_s \tright \infty$ 
corresponds to a spin-gapped state, in which case we find the single-electron 
hoppings $t$ and $t_{\perp}$ to be irrelevant (last column of 
\Tabref{table:phases}), in correspondence with previous calculations for a 
single array of coupled quantum wires \cite{Emery2000}. In contrast, if $K_s 
\tright 0$ we have a spin gapless state and all the couplings considered here 
are irrelevant\,---\,the system consists of decoupled LLs.

% ------------------------------------------------------------------------------
% FIGURE
% ------------------------------------------------------------------------------
\begin{figure}
\centering
\includegraphics[width=0.45\textwidth]{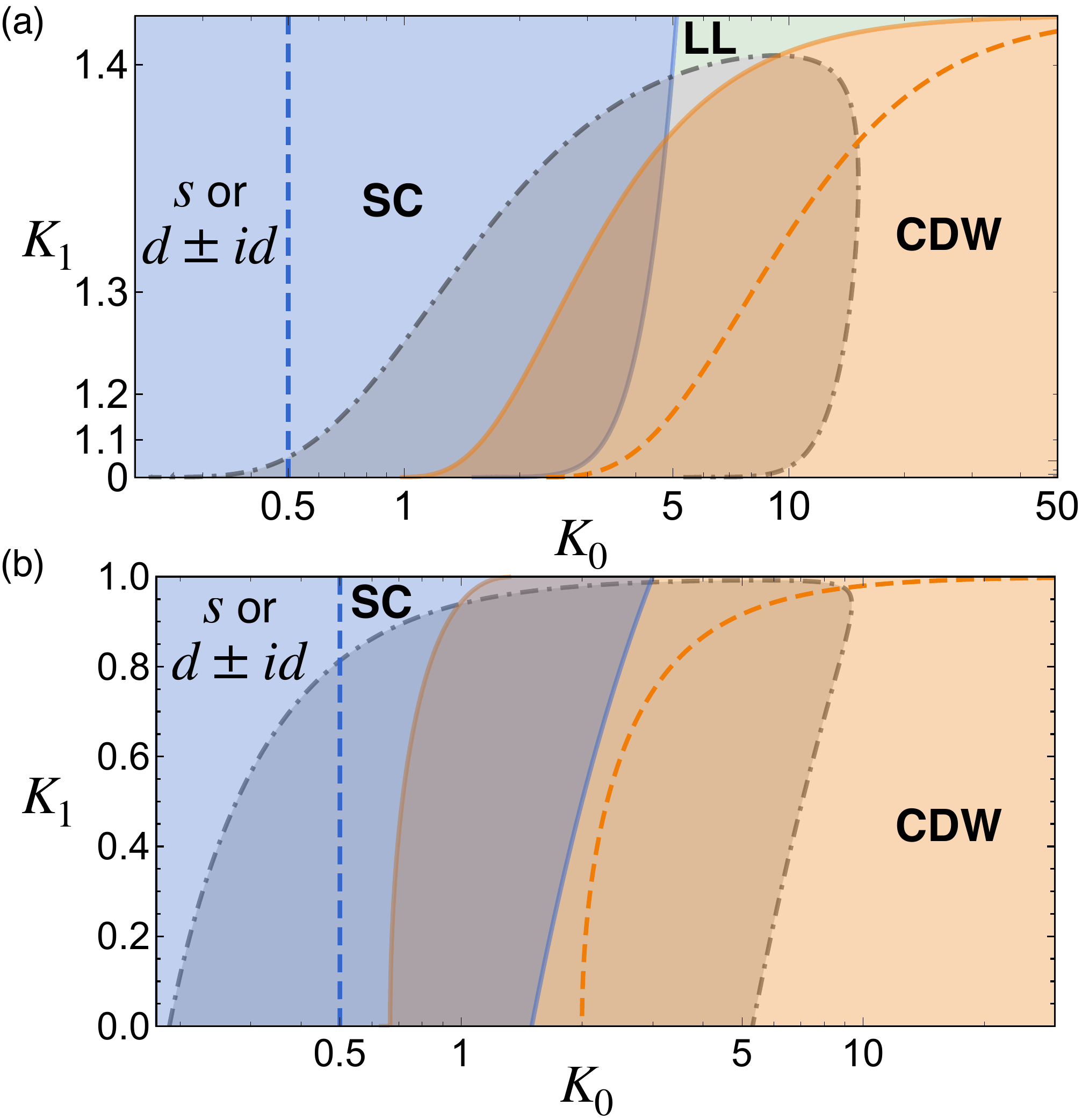}
\caption{
Phase diagram along the horizontal cuts marked in \Fref{fig:schematic} and for
$K_s \teq 2$: (a) $K_2 \teq 0.5$, (b) $K_2 \teq 0$. The parameters $K_0$ and 
$K_1$ (axes) are the Fourier coefficients defined in \Eqref{eq:kappa}.
In the region above the solid-blue line, one of $\mathcal{J}_1$ or 
$\mathcal{J}_2$ is relevant (SC order). In the region below the solid-orange 
line, at least one of  $\mathcal{V}_1$ and $\mathcal{V}_2$ is relevant (CDW 
order). The intra-array hopping ($t_\perp$) is relevant in the gray domain 
bounded by the dash-dotted line, implying that the system might be a Fermi 
liquid in this region. To the left of the blue-dashed line, the inter-array SC 
coupling is relevant, whereas the inter-array CDW coupling is relevant to the 
right of the orange-dashed line. The green area indicates a regime where all the 
couplings are irrelevant, corresponding to a 2D LL state. 
The main difference between (a) and (b) is the absence of the LL phase in the 
latter.
}
\label{fig:phase-diagram}
\end{figure}
% ------------------------------------------------------------------------------

% ------------------------------------------------------------------------------
\section{Phase diagram and analysis}
% ------------------------------------------------------------------------------

\subsection{Instability tendencies}

While one may explore any range of $K_s$, we will now focus on $K_s=2$. 
\Tabref{table:phases} shows that this falls in the regime where all couplings 
but $t$ are relevant and, therefore, it is representative of the physical 
scenarios involving phase competition, as is the case of MTBG, either driven by 
inter- or intra-array interactions (or both). Figure \ref{fig:phase-diagram} 
shows the phase diagram in two representative scenarios, defined by different 
magnitudes of the second harmonic in the Fourier expansion \eqref{eq:kappa}. 
Although $K_0$ is not strictly the Luttinger (charge) parameter of an individual 
wire, \Eqref{eq:action-array} implies it does represent the effective Luttinger 
parameter of an array behaving collectively as a LL \cite{Emery2000}. Therefore, 
$K_0 \tlt 1$ signals an effectively attractive regime while $K_0 \tgt 1$ 
describes repulsion. In this context, one qualitatively understands the fact 
that the SC phase (blue region) dominates in the small-$K_0$ portion of the 
phase diagram, while the CDW eventually becomes the only relevant phase for 
large $K_0$. 
In the crossover region $K_0 \tsim 1$, the domains of relevancy for the CDW and 
SC orders overlap; in addition, the intra-array hopping is relevant as well in 
this case (gray region enclosed by the dot-dashed line) which, should the 
hopping become dominant over the CDW and SC instabilities, implies the existence 
of a 2D FL phase. This indicates that the transition between SC and CDW with 
increasing repulsion (increasing $K_0$) can occur either directly or via an 
intervening FL phase, depending on the magnitude of $K_1$ (which is a measure of 
the nearest-neighbor inter-wire coupling within an array).
The precise outcome of this phase competition, or coexistence, depends on how 
the running of one coupling constant affects the others, whose analysis requires 
a perturbative RG calculation beyond first order, which is  not in the scope of 
this paper. 

When $K_2$ is finite, in addition to a FL, one finds that a LL phase is 
stabilized between the SC and CDW region close to the parameter-space boundary 
at $K_1 \tright \sqrt{2}$. This is marked by the green area in 
\Fref{fig:phase-diagram}(a) or the orange region in \Fref{fig:schematic}(b). 
Physically, the appearance of a LL phase in this case arises from the fact that 
$K_2$ promotes interaction between next-nearest-neighboring wires within an 
array, which is detrimental to the stability of the CDW. As first pointed out by 
Vishwanath and Carpentier \cite{Vishwanath2001}, when $K_2$ is included, near 
the $(K_1,K_2)$ space boundary, the minimum of the $\kappa(k_\perp)$ is located 
at some \emph{incommensurate} $k_\perp$ and the value is close to zero, which 
indicate strong fluctuations of a transverse incommensurate CDW order [the 
density correlation $\langle \phi^*_{c,k,k\perp} \phi_{c,k,k_\perp} \rangle 
\propto 1/\kappa(k_\perp) $]. The \emph{incommensurate} CDW fluctuations then 
destroy the crystallization so that all  couplings are irrelevant within that 
region. 
In contrast, for $K_2 \teq 0$, the most divergent transverse CDW is commensurate 
as $K_1 \tright 1$; i.e., $1/\kappa(k_{\perp}) \teq 1/K_0[1 \tplus 
\cos(k_{\perp} d)]$ diverges at $k_{\perp} \teq \pi/d$, so the 
next-nearest-neighbor intra-array CDW coupling will crystallize the system and 
there is no LL phase. Indeed, comparing \Frefs{fig:phase-diagram}(a) and 
\ref{fig:phase-diagram}(b), we see that a LL phase is stabilized at the expense 
of the CDW phase in the repulsive region ($K_0 \tgt 1$), without much impact on 
the region of SC stability. At large $K_0$, both intra-array (rightward off the 
solid orange line) and inter-array (rightward off the dashed orange line) CDW 
couplings are relevant\,---\,the electrons crystallize and an insulator ensues.

If $K_0 \tlt 0.5$, the inter-array Josephson coupling is relevant. At each wire 
crossing, the phases of the three SC order parameters couple via
\begin{equation}\label{eq:sc-coupling}
  \propto \mathcal{J}_0\sum_{i,j \in \{1,2,3\}} 
    \!\!\!\!\! \cos(\varphi^i_{\mathbf{r}} - \varphi^j_{\mathbf{r}}).
\end{equation}
Assuming that the intra-array SC coupling promotes uniform SC within each array, 
\Eqref{eq:sc-coupling} indicates that the global SC phase depends on the sign of 
$\mathcal{J}_0$: if $\mathcal{J}_0 \tlt 0$, the Josephson coupling favors 
$s$-wave SC with all $\varphi_i$ equal; but, if $\mathcal{J}_0 \tgt 0$, that 
coupling is frustrated and will result in a $2\pi/3$ difference between the 
phase of the SC order parameter of one array ($j$) with respect to the next 
($j+1$). This originates a $d \tpm id$ SC symmetry. A similar conclusion has 
been drawn by Wu \etal{} who have further considered triplet pairing and discuss 
the additional possibility of $p\tpm ip$ symmetry \cite{Wu2019}. 

% ------------------------------------------------------------------------------
\subsection{Commensurability}
% ------------------------------------------------------------------------------

In a conventional (i.e., single) LL problem, the proximity to commensurate 
electron densities is described by considering the Umklapp process within 
each wire \cite{Giamarchi2003} which, in the notation of \Eqref{eq:LL}, has the 
form 
\begin{equation}
  \frac{g_{U}}{(2\pi \alpha)^2} \int dx \cos[2\sqrt{2\pi}\phi_c +(4k_F-G)x],
\end{equation}
where $G$ is a vector of the reciprocal superlattice. The couplings $g_{U}$ and 
$K_c$ flow according to \Eqref{eq:sine-Gordon}, with the replacements $g_s \tto 
g_{U}$, $K_s \tto K_c$ \cite{Giamarchi2003}. In the present case, however, the 
effective Luttinger parameter $\kappa$ is a function of the transverse momentum 
[cf. \Eqref{eq:action-array}] due to the marginal interactions between wires 
within each array; this complicates the flow equations in the charge sector. We 
proceed by assuming, as a first approximation, that the flow equations for 
$g_{U}$ and $K_0$ behave analogously to those in \Eqref{eq:sine-Gordon}, in 
which case we naturally obtain distinct behavior at and away from half-filling: 
Our phase diagram in \Fref{fig:phase-diagram} indicates that, away from 
half-filling, the system is a SC provided $K_0$ is not too large; at (or near) 
half-filling, a large enough $g_{U}$ is able to drive the system to an 
insulating state even for very small $K_0$. Such SC-to-insulator transition is a 
general feature of the competing instabilities in a LL with commensurate 
density, because the Umklapp terms provide a ``condensation'' energy gain that 
ultimately makes the charge-gapped CDW state energetically favorable 
\cite{Giamarchi1997,Giamarchi2003,Emery1979}.

This competition between SC and CDW insulating states bears directly on the 
current experimental observations with magic-angle MTBG, which show the 
ground-state to be either a FL at generic densities, a SC near commensurate 
fillings, or an insulator at commensurability \cite{Cao2018, Cao2018a}\,---\,the 
coupled LL scenario is consistent with such observation.
For quantitative comparisons in this regard, it is worth noting that the 
electronic filling/density reported for the 2D experimental system needs to be 
converted to 1D electronic densities by taking into account that, in the 
coupled-wire picture, each Moir\'e unit cell contains three non-equivalent wires. 
For example, increasing the electron density by one electron per Moir\'e unit 
cell away from charge neutrality corresponds to adding $1/3$ electrons per 
segment of each non-equivalent wire within that unit cell.

% ------------------------------------------------------------------------------
\section{Discussion}
% ------------------------------------------------------------------------------

\subsection{Landscape of correlated states}

The propagation of interacting electrons along the quantum channels provided 
by the well defined AB-BA domain boundaries of small-angle MTBG 
[\Fref{fig:schematic}(a)] provides a natural low-energy picture for the 
emergence of competing SC and insulating states. At generic densities and 
moderate Luttinger parameter ($|K_0| \tsim 1$), we obtain SC and possibly 
FL as the dominant phases, with SC stabilized even for \emph{repulsive} Coulomb 
interactions ($K_0 \tgt 1$), which is noteworthy; at commensurate densities, 
the system is a charge-insulator. This holds both when $K_s \tlt 1$ 
and $K_s \tgt 1$, particularly in the spin-gapped regime ($K_s {\,\gg\,} 1$) 
where the only qualitative difference is the possible loss of the FL phase at 
very high $K_s$\,---\,this is significant for the model applicability to 
MTBG, where a magnetic field has been seen to destroy the insulating state 
\cite{Cao2018, Yankowitz2019}. 

\subsection{Anomalous metallic behavior}

Most interestingly, we see that the interaction among parallel wires contributes 
to stabilize both ``sliding'' and ``crossed sliding'' LL phases, thus extending 
previous findings for square arrays \cite{Emery2000, Vishwanath2001, 
Mukhopadhyay2001, Mukhopadhyay2001a} to this triangular geometry as well. These 
phases are extremely interesting because, on the one hand, they define a regime 
of metallic 2D transport underpinned entirely by Luttinger-liquid behavior and 
interactions, with the consequence that physical observables scale anomalously 
with temperature, size, and fields \cite{Mukhopadhyay2001}. On the other hand, 
and as a result, these regimes of 2D transport are entirely different from that 
of an effective circuit of independent 1D wires. 
Perhaps most significantly for current experiments is the fact that charge 
transport in these phases would have anisotropic fingerprints and an anomalous 
temperature dependence, thus being a \emph{natural} candidate for the ``strange 
metal'' behavior reported in magic-angle MTBG right above the temperatures 
where the insulating and SC states disappear \cite{Cao2019}. Moreover, in this 
picture, a symmetry breaking among the three equivalent wire arrays would 
naturally impart both local and global electronic ``nematicity'', a feature that 
has recently been inferred from high-resolution scanning tunneling 
microscopy (STM) experiments \cite{Jiang2019, Kerelsky2019, Choi2019a, Xie2019}.

\subsection{The nature of the 1D wire net}

As the presence of 1D modes traveling along the AB/BA domain boundaries is a 
decisive precondition for modeling MTBG in terms of the proposed network of 
coupled Luttinger liquids, in the remainder we elaborate on their evidence so 
far as well as on means through which they can be ensured.

It is now well established that MTBG undergoes considerable internal deformation 
within each Moir\'e unit cell so as to maximize extension of the energetically 
more favorable Bernal stacking at the expense of the AA-stacked regions. This 
energetic tendency is constrained by frustration at the interface between AB and 
BA regions and results in bilayers with uniform Bernal stacking essentially 
everywhere, except at sharp AB/BA domain walls and AA vertices whose geometry is 
depicted in \Fref{fig:schematic}(a). 
Numerical calculations reveal this relaxation effect very clearly 
\cite{Alden2013, Nam2017, Anelkovic2017, Gargiulo2018, Lucignano2019}: it 
starts becoming prominent for twist angles below $\tsim 2\text{--}3^\circ$ 
\cite{Wijk2015, Zhang2018b, Gargiulo2018} and is completely established for 
$\theta{\,\lesssim\,} 1^\circ$, at which point the width of the domain 
boundaries saturates at $\tsim 6\text{--}9$\,nm (becomes independent of twist 
amount for smaller angles) \cite{Alden2013, Yin2016, Gargiulo2018}. 
This threshold is confirmed experimentally \cite{Yoo2018} and MTBG samples have 
been reported with domain walls extending up to the micron scale while retaining 
their atomic-scale width \cite{Alden2013}. 

The electronic modes localized at the AB/BA domain boundaries of deliberately 
\emph{biased} bilayer graphene are expected to behave as perfect 1D quantum 
wires, so long as intervalley scattering remains unimportant \cite{Martin2008}; 
explicit bandstructure calculations have recently shown this to be indeed 
realized in relaxed MTBL \cite{Walet2019a,Carr2019b}. 
More importantly, this has been confirmed by measurements that probed 
electric transport along isolated AB/BA boundaries extending over several 
microns, which revealed the expected conductance quantization at $\tapprox 
4e^2/h$ and orders of magnitude enhancement of the mean-free path associated 
with these modes, when compared with that elsewhere in the sample \cite{Ju2015, 
Li2016d}. Their confinement to the domain walls has been confirmed directly by 
local STM and STS measurements \cite{Yin2016, Huang2018} as well as indirectly: 
(i) by the observation of Aharanov-Bohm oscillations in magnetotransport with 
spatial periods that correlate with closed paths along adjacent domain walls 
\cite{Rickhaus2018, Xu2019}; (ii) by the saturation of resistance near 
$h/(4e^2)$ \cite{Xu2019, Yoo2018} and metallic temperature dependence despite 
increases in interlayer bias (i.e., resistivity saturation and metallic 
temperature dependence with increasing bulk gap) \cite{Xu2019}; (iii) and by 
the local enhancement of infrared optical conductivity at the AB/BA interfaces, 
which is associated with the presence of the 1D modes \cite{Sunku2018}.

All of the above indicates that the coupled-wire model should provide an 
adequate description of biased MTBG, where the tunable bulk gap and topological 
character of the 1D states ensure the robustness of these electronic modes, in 
the regime where the Fermi level remains within the bulk gap. In the case of 
magic-angle MTBG, it would be interesting to experimentally investigate the fate 
of the correlated insulator and SC states under a finite interlayer bias. 

Whether these domain-wall-bound modes survive and remain influential at zero 
interlayer bias can depend on the conditions of the substrate. For example: 
these confined modes have been seen directly by STM/STS against a gapped Bernal 
background of graphene bilayers deposited on graphite, without any electrostatic 
bias \cite{Yin2016}; and it is know that, similarly to a graphene monolayer 
\cite{Kindermann2012, Hunt2013, Chen2014b, Moon2014a, Jung2015}, the Moir\'e and 
relaxation induced by boron nitride substrates generates a spectral gap for 
certain crystallographic orientations in bilayer graphene as well 
\cite{Mucha-Kruczynski2013, Moon2014a, Dean2013, Kim2018a}. Therefore, a bulk 
gap that guarantees and stabilizes the 1D modes can be engineered with 
appropriate substrate conditions. 

Finally, we note that there is a strong pseudomagnetic field due to the lattice 
relaxation, with magnitudes that might exceed 10\,T \cite{Wijk2015, Nam2017}. 
As a result of the triangular shape of the AB and BA domains, that field is 
quasi-uniform within the Bernal regions, but with opposite polarity\,---\,the 
polarity sharply switches precisely along the domain walls. The combined effect 
of large pseudomagnetic fields and abrupt polarity changes along the domain 
walls is likely to efficiently confine snake-type chiral states 
\cite{Jones:2017, Muller1992}. These would be chiral 1D modes of a different 
kind, which do not require a bulk gap \cite{Jones:2017}. 

\subsection{Conclusions}

Different experimental probes and theoretical work are persuasive enough of the 
conclusion that the triangular array of coupled quantum wires illustrated in 
\Fref{fig:schematic}(a) is the natural starting point to describe transport and 
correlated states in biased MTBG. The accuracy of this picture increases with 
larger bulk gaps and angles $\lesssim 1^{\circ}$, which ensures the sharpest 
domain boundaries as well as sufficiently long channels between the AA vertices 
for a valid LL description of each quantum wire (the inter-vertex distance is 
14\,nm for $\theta \teq 1^\circ$). The facts that such modes have been equally 
seen in unbiased experiments and that pseudomagnetic fields can themselves beget 
additional 1D states of a different nature, suggest the relevance of this 
description to unbiased devices as well. Indeed, the phenomenology of the 
correlated states, which so far has been scrutinized only in the unbiased case, 
tallies with the phase diagram arising from the coupled LL model, namely when it 
comes to: the types of correlated states involved and their competition, the 
sequence of phase transitions with doping, the association of CDW insulating 
states with commensurate fillings, the existence of non-Fermi liquid metallic 
states, and nematicity. 

Seeing as the detailed mechanisms underpinning both the insulating and SC states 
in MTBG remain an open problem, it is of utmost interest to experimentally 
scrutinize the evolution of the correlated phase diagram in MTBG (at the magic 
as well as smaller angles) \emph{as a function of the bulk gap} through 
interlayer bias. This would place the system in the regime where our model 
most reliably applies, while it would also assess its relevance to 
the strictly unbiased case. 

\emph{Note}\,---\,Recently, a preprint emerged with a 
similar formulation  \cite{Chou2019}, but with more restricted applicability 
directly to MTBG: it considers a wire net with $C_4$ symmetry rather than $C_6$, 
assumes persistent local SC order in puddle regions that encompass the wire 
intersections, and couplings are considered only at those intersections, without 
intra-array interactions. In specific cases (parameter ranges) where the two 
models can be compared, the conclusions agree.

\begin{acknowledgments}
CC thanks X.~Y.~Gu and J.~N.~Leaw for helpful discussions. This work was 
supported by the National Research Foundation of Singapore under its 
Medium-Sized Centre Programme.
\end{acknowledgments}

% ------------------------------------------------------------------------------
% REFERENCES
% ------------------------------------------------------------------------------
\bibliographystyle{apsrev4-1}
\bibliography{TBG}

\begin{thebibliography}{78}%
\makeatletter
\providecommand \@ifxundefined [1]{%
 \@ifx{#1\undefined}
}%
\providecommand \@ifnum [1]{%
 \ifnum #1\expandafter \@firstoftwo
 \else \expandafter \@secondoftwo
 \fi
}%
\providecommand \@ifx [1]{%
 \ifx #1\expandafter \@firstoftwo
 \else \expandafter \@secondoftwo
 \fi
}%
\providecommand \natexlab [1]{#1}%
\providecommand \enquote  [1]{``#1''}%
\providecommand \bibnamefont  [1]{#1}%
\providecommand \bibfnamefont [1]{#1}%
\providecommand \citenamefont [1]{#1}%
\providecommand \href@noop [0]{\@secondoftwo}%
\providecommand \href [0]{\begingroup \@sanitize@url \@href}%
\providecommand \@href[1]{\@@startlink{#1}\@@href}%
\providecommand \@@href[1]{\endgroup#1\@@endlink}%
\providecommand \@sanitize@url [0]{\catcode `\\12\catcode `\$12\catcode
  `\&12\catcode `\#12\catcode `\^12\catcode `\_12\catcode `\%12\relax}%
\providecommand \@@startlink[1]{}%
\providecommand \@@endlink[0]{}%
\providecommand \url  [0]{\begingroup\@sanitize@url \@url }%
\providecommand \@url [1]{\endgroup\@href {#1}{\urlprefix }}%
\providecommand \urlprefix  [0]{URL }%
\providecommand \Eprint [0]{\href }%
\providecommand \doibase [0]{http://dx.doi.org/}%
\providecommand \selectlanguage [0]{\@gobble}%
\providecommand \bibinfo  [0]{\@secondoftwo}%
\providecommand \bibfield  [0]{\@secondoftwo}%
\providecommand \translation [1]{[#1]}%
\providecommand \BibitemOpen [0]{}%
\providecommand \bibitemStop [0]{}%
\providecommand \bibitemNoStop [0]{.\EOS\space}%
\providecommand \EOS [0]{\spacefactor3000\relax}%
\providecommand \BibitemShut  [1]{\csname bibitem#1\endcsname}%
\let\auto@bib@innerbib\@empty
%</preamble>
\bibitem [{\citenamefont {Haldane}(1981)}]{Haldane1981}%
  \BibitemOpen
  \bibfield  {author} {\bibinfo {author} {\bibfnamefont {F.~D.~M.}\
  \bibnamefont {Haldane}},\ }\href {\doibase 10.1088/0022-3719/14/19/010}
  {\bibfield  {journal} {\bibinfo  {journal} {Journal of Physics C: Solid State
  Physics}\ }\textbf {\bibinfo {volume} {14}},\ \bibinfo {pages} {2585}
  (\bibinfo {year} {1981})}\BibitemShut {NoStop}%
\bibitem [{\citenamefont {Giamarchi}(2003)}]{Giamarchi2003}%
  \BibitemOpen
  \bibfield  {author} {\bibinfo {author} {\bibfnamefont {T.}~\bibnamefont
  {Giamarchi}},\ }\href {\doibase 10.1093/acprof:oso/9780198525004.001.0001}
  {\emph {\bibinfo {title} {Quantum Physics in One Dimension}}}\ (\bibinfo
  {publisher} {Oxford University Press},\ \bibinfo {year} {2003})\BibitemShut
  {NoStop}%
\bibitem [{\citenamefont {Fradkin}(2013)}]{Fradkin2013}%
  \BibitemOpen
  \bibfield  {author} {\bibinfo {author} {\bibfnamefont {E.}~\bibnamefont
  {Fradkin}},\ }\href {\doibase 10.1017/cbo9781139015509} {\emph {\bibinfo
  {title} {Field Theories of Condensed Matter Physics}}}\ (\bibinfo
  {publisher} {Cambridge University Press},\ \bibinfo {year}
  {2013})\BibitemShut {NoStop}%
\bibitem [{\citenamefont {Bednorz}\ and\ \citenamefont
  {M{\"u}ller}(1986)}]{Bednorz1986}%
  \BibitemOpen
  \bibfield  {author} {\bibinfo {author} {\bibfnamefont {J.~G.}\ \bibnamefont
  {Bednorz}}\ and\ \bibinfo {author} {\bibfnamefont {K.~A.}\ \bibnamefont
  {M{\"u}ller}},\ }\href {\doibase 10.1007/BF01303701} {\bibfield  {journal}
  {\bibinfo  {journal} {Zeitschrift f{\"u}r Physik B Condensed Matter}\
  }\textbf {\bibinfo {volume} {64}},\ \bibinfo {pages} {189} (\bibinfo {year}
  {1986})}\BibitemShut {NoStop}%
\bibitem [{\citenamefont {Tranquada}\ \emph {et~al.}(1994)\citenamefont
  {Tranquada}, \citenamefont {Buttrey}, \citenamefont {Sachan},\ and\
  \citenamefont {Lorenzo}}]{Tranquada1994}%
  \BibitemOpen
  \bibfield  {author} {\bibinfo {author} {\bibfnamefont {J.~M.}\ \bibnamefont
  {Tranquada}}, \bibinfo {author} {\bibfnamefont {D.~J.}\ \bibnamefont
  {Buttrey}}, \bibinfo {author} {\bibfnamefont {V.}~\bibnamefont {Sachan}}, \
  and\ \bibinfo {author} {\bibfnamefont {J.~E.}\ \bibnamefont {Lorenzo}},\
  }\href {\doibase 10.1103/PhysRevLett.73.1003} {\bibfield  {journal} {\bibinfo
   {journal} {Phys. Rev. Lett.}\ }\textbf {\bibinfo {volume} {73}},\ \bibinfo
  {pages} {1003} (\bibinfo {year} {1994})}\BibitemShut {NoStop}%
\bibitem [{\citenamefont {Zaanen}\ and\ \citenamefont
  {Gunnarsson}(1989)}]{Zaanen1989}%
  \BibitemOpen
  \bibfield  {author} {\bibinfo {author} {\bibfnamefont {J.}~\bibnamefont
  {Zaanen}}\ and\ \bibinfo {author} {\bibfnamefont {O.}~\bibnamefont
  {Gunnarsson}},\ }\href {\doibase 10.1103/PhysRevB.40.7391} {\bibfield
  {journal} {\bibinfo  {journal} {Phys. Rev. B}\ }\textbf {\bibinfo {volume}
  {40}},\ \bibinfo {pages} {7391} (\bibinfo {year} {1989})}\BibitemShut
  {NoStop}%
\bibitem [{\citenamefont {Machida}(1989)}]{Machida1989}%
  \BibitemOpen
  \bibfield  {author} {\bibinfo {author} {\bibfnamefont {K.}~\bibnamefont
  {Machida}},\ }\href {\doibase 10.1016/0921-4534(89)90316-X} {\bibfield
  {journal} {\bibinfo  {journal} {Phys. C Supercond.}\ }\textbf {\bibinfo
  {volume} {158}},\ \bibinfo {pages} {192} (\bibinfo {year}
  {1989})}\BibitemShut {NoStop}%
\bibitem [{\citenamefont {Kato}\ \emph {et~al.}(1990)\citenamefont {Kato},
  \citenamefont {Machida}, \citenamefont {Nakanishi},\ and\ \citenamefont
  {Fujita}}]{Kato1990}%
  \BibitemOpen
  \bibfield  {author} {\bibinfo {author} {\bibfnamefont {M.}~\bibnamefont
  {Kato}}, \bibinfo {author} {\bibfnamefont {K.}~\bibnamefont {Machida}},
  \bibinfo {author} {\bibfnamefont {H.}~\bibnamefont {Nakanishi}}, \ and\
  \bibinfo {author} {\bibfnamefont {M.}~\bibnamefont {Fujita}},\ }\href
  {\doibase 10.1143/JPSJ.59.1047} {\bibfield  {journal} {\bibinfo  {journal}
  {J. Phys. Soc. Japan}\ }\textbf {\bibinfo {volume} {59}},\ \bibinfo {pages}
  {1047} (\bibinfo {year} {1990})}\BibitemShut {NoStop}%
\bibitem [{\citenamefont {Emery}\ \emph {et~al.}(1999)\citenamefont {Emery},
  \citenamefont {Kivelson},\ and\ \citenamefont {Tranquada}}]{Emery1999}%
  \BibitemOpen
  \bibfield  {author} {\bibinfo {author} {\bibfnamefont {V.~J.}\ \bibnamefont
  {Emery}}, \bibinfo {author} {\bibfnamefont {S.~A.}\ \bibnamefont {Kivelson}},
  \ and\ \bibinfo {author} {\bibfnamefont {J.~M.}\ \bibnamefont {Tranquada}},\
  }\href {\doibase 10.1073/pnas.96.16.8814} {\bibfield  {journal} {\bibinfo
  {journal} {Proc. Natl. Acad. Sci.}\ }\textbf {\bibinfo {volume} {96}},\
  \bibinfo {pages} {8814} (\bibinfo {year} {1999})}\BibitemShut {NoStop}%
\bibitem [{\citenamefont {Strong}\ \emph {et~al.}(1994)\citenamefont {Strong},
  \citenamefont {Clarke},\ and\ \citenamefont {Anderson}}]{Strong1994}%
  \BibitemOpen
  \bibfield  {author} {\bibinfo {author} {\bibfnamefont {S.~P.}\ \bibnamefont
  {Strong}}, \bibinfo {author} {\bibfnamefont {D.~G.}\ \bibnamefont {Clarke}},
  \ and\ \bibinfo {author} {\bibfnamefont {P.~W.}\ \bibnamefont {Anderson}},\
  }\href {\doibase 10.1103/PhysRevLett.73.1007} {\bibfield  {journal} {\bibinfo
   {journal} {Phys. Rev. Lett.}\ }\textbf {\bibinfo {volume} {73}},\ \bibinfo
  {pages} {1007} (\bibinfo {year} {1994})}\BibitemShut {NoStop}%
\bibitem [{\citenamefont {Wen}(1990)}]{Wen1990}%
  \BibitemOpen
  \bibfield  {author} {\bibinfo {author} {\bibfnamefont {X.~G.}\ \bibnamefont
  {Wen}},\ }\href {\doibase 10.1103/PhysRevB.42.6623} {\bibfield  {journal}
  {\bibinfo  {journal} {Phys. Rev. B}\ }\textbf {\bibinfo {volume} {42}},\
  \bibinfo {pages} {6623} (\bibinfo {year} {1990})}\BibitemShut {NoStop}%
\bibitem [{\citenamefont {Emery}\ \emph {et~al.}(2000)\citenamefont {Emery},
  \citenamefont {Fradkin}, \citenamefont {Kivelson},\ and\ \citenamefont
  {Lubensky}}]{Emery2000}%
  \BibitemOpen
  \bibfield  {author} {\bibinfo {author} {\bibfnamefont {V.~J.}\ \bibnamefont
  {Emery}}, \bibinfo {author} {\bibfnamefont {E.}~\bibnamefont {Fradkin}},
  \bibinfo {author} {\bibfnamefont {S.~A.}\ \bibnamefont {Kivelson}}, \ and\
  \bibinfo {author} {\bibfnamefont {T.~C.}\ \bibnamefont {Lubensky}},\ }\href
  {\doibase 10.1103/PhysRevLett.85.2160} {\bibfield  {journal} {\bibinfo
  {journal} {Phys. Rev. Lett.}\ }\textbf {\bibinfo {volume} {85}},\ \bibinfo
  {pages} {2160} (\bibinfo {year} {2000})}\BibitemShut {NoStop}%
\bibitem [{\citenamefont {Vishwanath}\ and\ \citenamefont
  {Carpentier}(2001)}]{Vishwanath2001}%
  \BibitemOpen
  \bibfield  {author} {\bibinfo {author} {\bibfnamefont {A.}~\bibnamefont
  {Vishwanath}}\ and\ \bibinfo {author} {\bibfnamefont {D.}~\bibnamefont
  {Carpentier}},\ }\href {\doibase 10.1103/PhysRevLett.86.676} {\bibfield
  {journal} {\bibinfo  {journal} {Phys. Rev. Lett.}\ }\textbf {\bibinfo
  {volume} {86}},\ \bibinfo {pages} {676} (\bibinfo {year} {2001})}\BibitemShut
  {NoStop}%
\bibitem [{\citenamefont {Mukhopadhyay}\ \emph
  {et~al.}(2001{\natexlab{a}})\citenamefont {Mukhopadhyay}, \citenamefont
  {Kane},\ and\ \citenamefont {Lubensky}}]{Mukhopadhyay2001}%
  \BibitemOpen
  \bibfield  {author} {\bibinfo {author} {\bibfnamefont {R.}~\bibnamefont
  {Mukhopadhyay}}, \bibinfo {author} {\bibfnamefont {C.~L.}\ \bibnamefont
  {Kane}}, \ and\ \bibinfo {author} {\bibfnamefont {T.~C.}\ \bibnamefont
  {Lubensky}},\ }\href {\doibase 10.1103/PhysRevB.63.081103} {\bibfield
  {journal} {\bibinfo  {journal} {Phys. Rev. B}\ }\textbf {\bibinfo {volume}
  {63}},\ \bibinfo {pages} {081103} (\bibinfo {year}
  {2001}{\natexlab{a}})}\BibitemShut {NoStop}%
\bibitem [{\citenamefont {Mukhopadhyay}\ \emph
  {et~al.}(2001{\natexlab{b}})\citenamefont {Mukhopadhyay}, \citenamefont
  {Kane},\ and\ \citenamefont {Lubensky}}]{Mukhopadhyay2001a}%
  \BibitemOpen
  \bibfield  {author} {\bibinfo {author} {\bibfnamefont {R.}~\bibnamefont
  {Mukhopadhyay}}, \bibinfo {author} {\bibfnamefont {C.~L.}\ \bibnamefont
  {Kane}}, \ and\ \bibinfo {author} {\bibfnamefont {T.~C.}\ \bibnamefont
  {Lubensky}},\ }\href {\doibase 10.1103/PhysRevB.64.045120} {\bibfield
  {journal} {\bibinfo  {journal} {Phys. Rev. B}\ }\textbf {\bibinfo {volume}
  {64}},\ \bibinfo {pages} {045120} (\bibinfo {year}
  {2001}{\natexlab{b}})}\BibitemShut {NoStop}%
\bibitem [{\citenamefont {Cao}\ \emph {et~al.}(2018{\natexlab{a}})\citenamefont
  {Cao}, \citenamefont {Fatemi}, \citenamefont {Demir}, \citenamefont {Fang},
  \citenamefont {Tomarken}, \citenamefont {Luo}, \citenamefont
  {Sanchez-Yamagishi}, \citenamefont {Watanabe}, \citenamefont {Taniguchi},
  \citenamefont {Kaxiras}, \citenamefont {Ashoori},\ and\ \citenamefont
  {Jarillo-Herrero}}]{Cao2018}%
  \BibitemOpen
  \bibfield  {author} {\bibinfo {author} {\bibfnamefont {Y.}~\bibnamefont
  {Cao}}, \bibinfo {author} {\bibfnamefont {V.}~\bibnamefont {Fatemi}},
  \bibinfo {author} {\bibfnamefont {A.}~\bibnamefont {Demir}}, \bibinfo
  {author} {\bibfnamefont {S.}~\bibnamefont {Fang}}, \bibinfo {author}
  {\bibfnamefont {S.~L.}\ \bibnamefont {Tomarken}}, \bibinfo {author}
  {\bibfnamefont {J.~Y.}\ \bibnamefont {Luo}}, \bibinfo {author} {\bibfnamefont
  {J.~D.}\ \bibnamefont {Sanchez-Yamagishi}}, \bibinfo {author} {\bibfnamefont
  {K.}~\bibnamefont {Watanabe}}, \bibinfo {author} {\bibfnamefont
  {T.}~\bibnamefont {Taniguchi}}, \bibinfo {author} {\bibfnamefont
  {E.}~\bibnamefont {Kaxiras}}, \bibinfo {author} {\bibfnamefont {R.~C.}\
  \bibnamefont {Ashoori}}, \ and\ \bibinfo {author} {\bibfnamefont
  {P.}~\bibnamefont {Jarillo-Herrero}},\ }\href {\doibase 10.1038/nature26154}
  {\bibfield  {journal} {\bibinfo  {journal} {Nature}\ }\textbf {\bibinfo
  {volume} {556}},\ \bibinfo {pages} {80} (\bibinfo {year}
  {2018}{\natexlab{a}})}\BibitemShut {NoStop}%
\bibitem [{\citenamefont {Cao}\ \emph {et~al.}(2018{\natexlab{b}})\citenamefont
  {Cao}, \citenamefont {Fatemi}, \citenamefont {Fang}, \citenamefont
  {Watanabe}, \citenamefont {Taniguchi}, \citenamefont {Kaxiras},\ and\
  \citenamefont {Jarillo-Herrero}}]{Cao2018a}%
  \BibitemOpen
  \bibfield  {author} {\bibinfo {author} {\bibfnamefont {Y.}~\bibnamefont
  {Cao}}, \bibinfo {author} {\bibfnamefont {V.}~\bibnamefont {Fatemi}},
  \bibinfo {author} {\bibfnamefont {S.}~\bibnamefont {Fang}}, \bibinfo {author}
  {\bibfnamefont {K.}~\bibnamefont {Watanabe}}, \bibinfo {author}
  {\bibfnamefont {T.}~\bibnamefont {Taniguchi}}, \bibinfo {author}
  {\bibfnamefont {E.}~\bibnamefont {Kaxiras}}, \ and\ \bibinfo {author}
  {\bibfnamefont {P.}~\bibnamefont {Jarillo-Herrero}},\ }\href {\doibase
  10.1038/nature26160} {\bibfield  {journal} {\bibinfo  {journal} {Nature}\
  }\textbf {\bibinfo {volume} {556}},\ \bibinfo {pages} {43} (\bibinfo {year}
  {2018}{\natexlab{b}})}\BibitemShut {NoStop}%
\bibitem [{\citenamefont {{Su{\'{a}}rez Morell}}\ \emph
  {et~al.}(2010)\citenamefont {{Su{\'{a}}rez Morell}}, \citenamefont {Correa},
  \citenamefont {Vargas}, \citenamefont {Pacheco},\ and\ \citenamefont
  {Barticevic}}]{SuarezMorell2010}%
  \BibitemOpen
  \bibfield  {author} {\bibinfo {author} {\bibfnamefont {E.}~\bibnamefont
  {{Su{\'{a}}rez Morell}}}, \bibinfo {author} {\bibfnamefont {J.~D.}\
  \bibnamefont {Correa}}, \bibinfo {author} {\bibfnamefont {P.}~\bibnamefont
  {Vargas}}, \bibinfo {author} {\bibfnamefont {M.}~\bibnamefont {Pacheco}}, \
  and\ \bibinfo {author} {\bibfnamefont {Z.}~\bibnamefont {Barticevic}},\
  }\href {\doibase 10.1103/PhysRevB.82.121407} {\bibfield  {journal} {\bibinfo
  {journal} {Phys. Rev. B}\ }\textbf {\bibinfo {volume} {82}},\ \bibinfo
  {pages} {121407} (\bibinfo {year} {2010})}\BibitemShut {NoStop}%
\bibitem [{\citenamefont {Li}\ \emph {et~al.}(2010)\citenamefont {Li},
  \citenamefont {Luican}, \citenamefont {{Lopes dos Santos}}, \citenamefont
  {{Castro Neto}}, \citenamefont {Reina}, \citenamefont {Kong},\ and\
  \citenamefont {Andrei}}]{Li2010}%
  \BibitemOpen
  \bibfield  {author} {\bibinfo {author} {\bibfnamefont {G.}~\bibnamefont
  {Li}}, \bibinfo {author} {\bibfnamefont {A.}~\bibnamefont {Luican}}, \bibinfo
  {author} {\bibfnamefont {J.~M.~B.}\ \bibnamefont {{Lopes dos Santos}}},
  \bibinfo {author} {\bibfnamefont {A.~H.}\ \bibnamefont {{Castro Neto}}},
  \bibinfo {author} {\bibfnamefont {A.}~\bibnamefont {Reina}}, \bibinfo
  {author} {\bibfnamefont {J.}~\bibnamefont {Kong}}, \ and\ \bibinfo {author}
  {\bibfnamefont {E.~Y.}\ \bibnamefont {Andrei}},\ }\href {\doibase
  10.1038/nphys1463} {\bibfield  {journal} {\bibinfo  {journal} {Nat. Phys.}\
  }\textbf {\bibinfo {volume} {6}},\ \bibinfo {pages} {109} (\bibinfo {year}
  {2010})}\BibitemShut {NoStop}%
\bibitem [{\citenamefont {Bistritzer}\ and\ \citenamefont
  {MacDonald}(2011)}]{Bistritzer2011}%
  \BibitemOpen
  \bibfield  {author} {\bibinfo {author} {\bibfnamefont {R.}~\bibnamefont
  {Bistritzer}}\ and\ \bibinfo {author} {\bibfnamefont {A.~H.}\ \bibnamefont
  {MacDonald}},\ }\href {\doibase 10.1073/pnas.1108174108} {\bibfield
  {journal} {\bibinfo  {journal} {Proc. Natl. Acad. Sci.}\ }\textbf {\bibinfo
  {volume} {108}},\ \bibinfo {pages} {12233} (\bibinfo {year}
  {2011})}\BibitemShut {NoStop}%
\bibitem [{\citenamefont {{Lopes dos Santos}}\ \emph
  {et~al.}(2012)\citenamefont {{Lopes dos Santos}}, \citenamefont {Peres},\
  and\ \citenamefont {{Castro Neto}}}]{LopesdosSantos2012}%
  \BibitemOpen
  \bibfield  {author} {\bibinfo {author} {\bibfnamefont {J.~M.~B.}\
  \bibnamefont {{Lopes dos Santos}}}, \bibinfo {author} {\bibfnamefont
  {N.~M.~R.}\ \bibnamefont {Peres}}, \ and\ \bibinfo {author} {\bibfnamefont
  {A.~H.}\ \bibnamefont {{Castro Neto}}},\ }\href {\doibase
  10.1103/PhysRevB.86.155449} {\bibfield  {journal} {\bibinfo  {journal} {Phys.
  Rev. B}\ }\textbf {\bibinfo {volume} {86}},\ \bibinfo {pages} {155449}
  (\bibinfo {year} {2012})}\BibitemShut {NoStop}%
\bibitem [{\citenamefont {{Lopes dos Santos}}\ \emph
  {et~al.}(2007)\citenamefont {{Lopes dos Santos}}, \citenamefont {Peres},\
  and\ \citenamefont {{Castro Neto}}}]{LopesdosSantos2007}%
  \BibitemOpen
  \bibfield  {author} {\bibinfo {author} {\bibfnamefont {J.~M.~B.}\
  \bibnamefont {{Lopes dos Santos}}}, \bibinfo {author} {\bibfnamefont
  {N.~M.~R.}\ \bibnamefont {Peres}}, \ and\ \bibinfo {author} {\bibfnamefont
  {A.~H.}\ \bibnamefont {{Castro Neto}}},\ }\href {\doibase
  10.1103/PhysRevLett.99.256802} {\bibfield  {journal} {\bibinfo  {journal}
  {Phys. Rev. Lett.}\ }\textbf {\bibinfo {volume} {99}},\ \bibinfo {pages}
  {256802} (\bibinfo {year} {2007})}\BibitemShut {NoStop}%
\bibitem [{\citenamefont {Koshino}\ \emph {et~al.}(2018)\citenamefont
  {Koshino}, \citenamefont {Yuan}, \citenamefont {Koretsune}, \citenamefont
  {Ochi}, \citenamefont {Kuroki},\ and\ \citenamefont {Fu}}]{Koshino2018}%
  \BibitemOpen
  \bibfield  {author} {\bibinfo {author} {\bibfnamefont {M.}~\bibnamefont
  {Koshino}}, \bibinfo {author} {\bibfnamefont {N.~F.~Q.}\ \bibnamefont
  {Yuan}}, \bibinfo {author} {\bibfnamefont {T.}~\bibnamefont {Koretsune}},
  \bibinfo {author} {\bibfnamefont {M.}~\bibnamefont {Ochi}}, \bibinfo {author}
  {\bibfnamefont {K.}~\bibnamefont {Kuroki}}, \ and\ \bibinfo {author}
  {\bibfnamefont {L.}~\bibnamefont {Fu}},\ }\href {\doibase
  10.1103/PhysRevX.8.031087} {\bibfield  {journal} {\bibinfo  {journal} {Phys.
  Rev. X}\ }\textbf {\bibinfo {volume} {8}},\ \bibinfo {pages} {031087}
  (\bibinfo {year} {2018})}\BibitemShut {NoStop}%
\bibitem [{\citenamefont {Po}\ \emph {et~al.}(2018)\citenamefont {Po},
  \citenamefont {Zou}, \citenamefont {Vishwanath},\ and\ \citenamefont
  {Senthil}}]{Po2018}%
  \BibitemOpen
  \bibfield  {author} {\bibinfo {author} {\bibfnamefont {H.~C.}\ \bibnamefont
  {Po}}, \bibinfo {author} {\bibfnamefont {L.}~\bibnamefont {Zou}}, \bibinfo
  {author} {\bibfnamefont {A.}~\bibnamefont {Vishwanath}}, \ and\ \bibinfo
  {author} {\bibfnamefont {T.}~\bibnamefont {Senthil}},\ }\href {\doibase
  10.1103/PhysRevX.8.031089} {\bibfield  {journal} {\bibinfo  {journal} {Phys.
  Rev. X}\ }\textbf {\bibinfo {volume} {8}},\ \bibinfo {pages} {031089}
  (\bibinfo {year} {2018})}\BibitemShut {NoStop}%
\bibitem [{\citenamefont {Kang}\ and\ \citenamefont {Vafek}(2018)}]{Kang2018}%
  \BibitemOpen
  \bibfield  {author} {\bibinfo {author} {\bibfnamefont {J.}~\bibnamefont
  {Kang}}\ and\ \bibinfo {author} {\bibfnamefont {O.}~\bibnamefont {Vafek}},\
  }\href {\doibase 10.1103/PhysRevX.8.031088} {\bibfield  {journal} {\bibinfo
  {journal} {Phys. Rev. X}\ }\textbf {\bibinfo {volume} {8}},\ \bibinfo {pages}
  {031088} (\bibinfo {year} {2018})}\BibitemShut {NoStop}%
\bibitem [{\citenamefont {Liu}\ \emph {et~al.}(2018)\citenamefont {Liu},
  \citenamefont {Zhang}, \citenamefont {Chen},\ and\ \citenamefont
  {Yang}}]{Liu2018}%
  \BibitemOpen
  \bibfield  {author} {\bibinfo {author} {\bibfnamefont {C.-C.}\ \bibnamefont
  {Liu}}, \bibinfo {author} {\bibfnamefont {L.-D.}\ \bibnamefont {Zhang}},
  \bibinfo {author} {\bibfnamefont {W.-Q.}\ \bibnamefont {Chen}}, \ and\
  \bibinfo {author} {\bibfnamefont {F.}~\bibnamefont {Yang}},\ }\href {\doibase
  10.1103/PhysRevLett.121.217001} {\bibfield  {journal} {\bibinfo  {journal}
  {Phys. Rev. Lett.}\ }\textbf {\bibinfo {volume} {121}},\ \bibinfo {pages}
  {217001} (\bibinfo {year} {2018})}\BibitemShut {NoStop}%
\bibitem [{\citenamefont {Isobe}\ \emph {et~al.}(2018)\citenamefont {Isobe},
  \citenamefont {Yuan},\ and\ \citenamefont {Fu}}]{Isobe2018}%
  \BibitemOpen
  \bibfield  {author} {\bibinfo {author} {\bibfnamefont {H.}~\bibnamefont
  {Isobe}}, \bibinfo {author} {\bibfnamefont {N.~F.~Q.}\ \bibnamefont {Yuan}},
  \ and\ \bibinfo {author} {\bibfnamefont {L.}~\bibnamefont {Fu}},\ }\href
  {\doibase 10.1103/PhysRevX.8.041041} {\bibfield  {journal} {\bibinfo
  {journal} {Phys. Rev. X}\ }\textbf {\bibinfo {volume} {8}},\ \bibinfo {pages}
  {041041} (\bibinfo {year} {2018})}\BibitemShut {NoStop}%
\bibitem [{\citenamefont {You}\ and\ \citenamefont
  {Vishwanath}(2019)}]{You2019}%
  \BibitemOpen
  \bibfield  {author} {\bibinfo {author} {\bibfnamefont {Y.-Z.}\ \bibnamefont
  {You}}\ and\ \bibinfo {author} {\bibfnamefont {A.}~\bibnamefont
  {Vishwanath}},\ }\href {\doibase 10.1038/s41535-019-0153-4} {\bibfield
  {journal} {\bibinfo  {journal} {npj Quantum Materials}\ }\textbf {\bibinfo
  {volume} {4}},\ \bibinfo {pages} {16} (\bibinfo {year} {2019})}\BibitemShut
  {NoStop}%
\bibitem [{\citenamefont {Laksono}\ \emph {et~al.}(2018)\citenamefont
  {Laksono}, \citenamefont {Leaw}, \citenamefont {Reaves}, \citenamefont
  {Singh}, \citenamefont {Wang}, \citenamefont {Adam},\ and\ \citenamefont
  {Gu}}]{Laksono2018}%
  \BibitemOpen
  \bibfield  {author} {\bibinfo {author} {\bibfnamefont {E.}~\bibnamefont
  {Laksono}}, \bibinfo {author} {\bibfnamefont {J.~N.}\ \bibnamefont {Leaw}},
  \bibinfo {author} {\bibfnamefont {A.}~\bibnamefont {Reaves}}, \bibinfo
  {author} {\bibfnamefont {M.}~\bibnamefont {Singh}}, \bibinfo {author}
  {\bibfnamefont {X.}~\bibnamefont {Wang}}, \bibinfo {author} {\bibfnamefont
  {S.}~\bibnamefont {Adam}}, \ and\ \bibinfo {author} {\bibfnamefont
  {X.}~\bibnamefont {Gu}},\ }\href {\doibase
  https://doi.org/10.1016/j.ssc.2018.07.013} {\bibfield  {journal} {\bibinfo
  {journal} {Solid State Commun.}\ }\textbf {\bibinfo {volume} {282}},\
  \bibinfo {pages} {38 } (\bibinfo {year} {2018})}\BibitemShut {NoStop}%
\bibitem [{\citenamefont {Xu}\ and\ \citenamefont {Balents}(2018)}]{Xu2018}%
  \BibitemOpen
  \bibfield  {author} {\bibinfo {author} {\bibfnamefont {C.}~\bibnamefont
  {Xu}}\ and\ \bibinfo {author} {\bibfnamefont {L.}~\bibnamefont {Balents}},\
  }\href {\doibase 10.1103/PhysRevLett.121.087001} {\bibfield  {journal}
  {\bibinfo  {journal} {Phys. Rev. Lett.}\ }\textbf {\bibinfo {volume} {121}},\
  \bibinfo {pages} {087001} (\bibinfo {year} {2018})}\BibitemShut {NoStop}%
\bibitem [{\citenamefont {Xu}\ \emph {et~al.}(2018)\citenamefont {Xu},
  \citenamefont {Law},\ and\ \citenamefont {Lee}}]{Xiao2018}%
  \BibitemOpen
  \bibfield  {author} {\bibinfo {author} {\bibfnamefont {X.~Y.}\ \bibnamefont
  {Xu}}, \bibinfo {author} {\bibfnamefont {K.~T.}\ \bibnamefont {Law}}, \ and\
  \bibinfo {author} {\bibfnamefont {P.~A.}\ \bibnamefont {Lee}},\ }\href
  {\doibase 10.1103/PhysRevB.98.121406} {\bibfield  {journal} {\bibinfo
  {journal} {Phys. Rev. B}\ }\textbf {\bibinfo {volume} {98}},\ \bibinfo
  {pages} {121406} (\bibinfo {year} {2018})}\BibitemShut {NoStop}%
\bibitem [{\citenamefont {{Gu}}\ \emph {et~al.}(2019)\citenamefont {{Gu}},
  \citenamefont {{Chen}}, \citenamefont {{Leaw}}, \citenamefont {{Laksono}},
  \citenamefont {{Pereira}}, \citenamefont {{Vignale}},\ and\ \citenamefont
  {{Adam}}}]{Gu2019}%
  \BibitemOpen
  \bibfield  {author} {\bibinfo {author} {\bibfnamefont {X.}~\bibnamefont
  {{Gu}}}, \bibinfo {author} {\bibfnamefont {C.}~\bibnamefont {{Chen}}},
  \bibinfo {author} {\bibfnamefont {J.~N.}\ \bibnamefont {{Leaw}}}, \bibinfo
  {author} {\bibfnamefont {E.}~\bibnamefont {{Laksono}}}, \bibinfo {author}
  {\bibfnamefont {V.~M.}\ \bibnamefont {{Pereira}}}, \bibinfo {author}
  {\bibfnamefont {G.}~\bibnamefont {{Vignale}}}, \ and\ \bibinfo {author}
  {\bibfnamefont {S.}~\bibnamefont {{Adam}}},\ }\href@noop {} {\bibfield
  {journal} {\bibinfo  {journal} {arXiv e-prints}\ ,\ \bibinfo {eid}
  {arXiv:1902.00029}} (\bibinfo {year} {2019})},\ \Eprint
  {http://arxiv.org/abs/1902.00029} {arXiv:1902.00029 [cond-mat.supr-con]}
  \BibitemShut {NoStop}%
\bibitem [{\citenamefont {Padhi}\ \emph {et~al.}(2018)\citenamefont {Padhi},
  \citenamefont {Setty},\ and\ \citenamefont {Phillips}}]{Padhi2018}%
  \BibitemOpen
  \bibfield  {author} {\bibinfo {author} {\bibfnamefont {B.}~\bibnamefont
  {Padhi}}, \bibinfo {author} {\bibfnamefont {C.}~\bibnamefont {Setty}}, \ and\
  \bibinfo {author} {\bibfnamefont {P.~W.}\ \bibnamefont {Phillips}},\ }\href
  {\doibase 10.1021/acs.nanolett.8b02033} {\bibfield  {journal} {\bibinfo
  {journal} {Nano Lett.}\ }\textbf {\bibinfo {volume} {18}},\ \bibinfo {pages}
  {6175} (\bibinfo {year} {2018})}\BibitemShut {NoStop}%
\bibitem [{\citenamefont {Padhi}\ and\ \citenamefont
  {Phillips}(2019)}]{Padhi2019}%
  \BibitemOpen
  \bibfield  {author} {\bibinfo {author} {\bibfnamefont {B.}~\bibnamefont
  {Padhi}}\ and\ \bibinfo {author} {\bibfnamefont {P.~W.}\ \bibnamefont
  {Phillips}},\ }\href {\doibase 10.1103/PhysRevB.99.205141} {\bibfield
  {journal} {\bibinfo  {journal} {Phys. Rev. B}\ }\textbf {\bibinfo {volume}
  {99}},\ \bibinfo {pages} {205141} (\bibinfo {year} {2019})}\BibitemShut
  {NoStop}%
\bibitem [{\citenamefont {Carr}\ \emph {et~al.}(2019)\citenamefont {Carr},
  \citenamefont {Fang}, \citenamefont {Zhu},\ and\ \citenamefont
  {Kaxiras}}]{Carr2019b}%
  \BibitemOpen
  \bibfield  {author} {\bibinfo {author} {\bibfnamefont {S.}~\bibnamefont
  {Carr}}, \bibinfo {author} {\bibfnamefont {S.}~\bibnamefont {Fang}}, \bibinfo
  {author} {\bibfnamefont {Z.}~\bibnamefont {Zhu}}, \ and\ \bibinfo {author}
  {\bibfnamefont {E.}~\bibnamefont {Kaxiras}},\ }\href {\doibase
  10.1103/PhysRevResearch.1.013001} {\bibfield  {journal} {\bibinfo  {journal}
  {Phys. Rev. Res.}\ }\textbf {\bibinfo {volume} {1}},\ \bibinfo {pages}
  {013001} (\bibinfo {year} {2019})}\BibitemShut {NoStop}%
\bibitem [{\citenamefont {Martin}\ \emph {et~al.}(2008)\citenamefont {Martin},
  \citenamefont {Blanter},\ and\ \citenamefont {Morpurgo}}]{Martin2008}%
  \BibitemOpen
  \bibfield  {author} {\bibinfo {author} {\bibfnamefont {I.}~\bibnamefont
  {Martin}}, \bibinfo {author} {\bibfnamefont {Y.~M.}\ \bibnamefont {Blanter}},
  \ and\ \bibinfo {author} {\bibfnamefont {A.~F.}\ \bibnamefont {Morpurgo}},\
  }\href {\doibase 10.1103/PhysRevLett.100.036804} {\bibfield  {journal}
  {\bibinfo  {journal} {Phys. Rev. Lett.}\ }\textbf {\bibinfo {volume} {100}},\
  \bibinfo {pages} {036804} (\bibinfo {year} {2008})}\BibitemShut {NoStop}%
\bibitem [{\citenamefont {Kindermann}\ \emph {et~al.}(2012)\citenamefont
  {Kindermann}, \citenamefont {Uchoa},\ and\ \citenamefont
  {Miller}}]{Kindermann2012}%
  \BibitemOpen
  \bibfield  {author} {\bibinfo {author} {\bibfnamefont {M.}~\bibnamefont
  {Kindermann}}, \bibinfo {author} {\bibfnamefont {B.}~\bibnamefont {Uchoa}}, \
  and\ \bibinfo {author} {\bibfnamefont {D.~L.}\ \bibnamefont {Miller}},\
  }\href {\doibase 10.1103/PhysRevB.86.115415} {\bibfield  {journal} {\bibinfo
  {journal} {Phys. Rev. B}\ }\textbf {\bibinfo {volume} {86}},\ \bibinfo
  {pages} {115415} (\bibinfo {year} {2012})}\BibitemShut {NoStop}%
\bibitem [{\citenamefont {Zhang}\ \emph {et~al.}(2013)\citenamefont {Zhang},
  \citenamefont {MacDonald},\ and\ \citenamefont {Mele}}]{Zhang2013}%
  \BibitemOpen
  \bibfield  {author} {\bibinfo {author} {\bibfnamefont {F.}~\bibnamefont
  {Zhang}}, \bibinfo {author} {\bibfnamefont {A.~H.}\ \bibnamefont
  {MacDonald}}, \ and\ \bibinfo {author} {\bibfnamefont {E.~J.}\ \bibnamefont
  {Mele}},\ }\href {\doibase 10.1073/pnas.1308853110} {\bibfield  {journal}
  {\bibinfo  {journal} {Proc. Natl. Acad. Sci.}\ }\textbf {\bibinfo {volume}
  {110}},\ \bibinfo {pages} {10546} (\bibinfo {year} {2013})}\BibitemShut
  {NoStop}%
\bibitem [{\citenamefont {Vaezi}\ \emph {et~al.}(2013)\citenamefont {Vaezi},
  \citenamefont {Liang}, \citenamefont {Ngai}, \citenamefont {Yang},\ and\
  \citenamefont {Kim}}]{Vaezi2013}%
  \BibitemOpen
  \bibfield  {author} {\bibinfo {author} {\bibfnamefont {A.}~\bibnamefont
  {Vaezi}}, \bibinfo {author} {\bibfnamefont {Y.}~\bibnamefont {Liang}},
  \bibinfo {author} {\bibfnamefont {D.~H.}\ \bibnamefont {Ngai}}, \bibinfo
  {author} {\bibfnamefont {L.}~\bibnamefont {Yang}}, \ and\ \bibinfo {author}
  {\bibfnamefont {E.-A.}\ \bibnamefont {Kim}},\ }\href {\doibase
  10.1103/PhysRevX.3.021018} {\bibfield  {journal} {\bibinfo  {journal} {Phys.
  Rev. X}\ }\textbf {\bibinfo {volume} {3}},\ \bibinfo {pages} {021018}
  (\bibinfo {year} {2013})}\BibitemShut {NoStop}%
\bibitem [{\citenamefont {Koshino}(2013)}]{Koshino2013}%
  \BibitemOpen
  \bibfield  {author} {\bibinfo {author} {\bibfnamefont {M.}~\bibnamefont
  {Koshino}},\ }\href {\doibase 10.1103/PhysRevB.88.115409} {\bibfield
  {journal} {\bibinfo  {journal} {Phys. Rev. B}\ }\textbf {\bibinfo {volume}
  {88}},\ \bibinfo {pages} {115409} (\bibinfo {year} {2013})}\BibitemShut
  {NoStop}%
\bibitem [{\citenamefont {San-Jose}\ and\ \citenamefont
  {Prada}(2013)}]{San-Jose2013}%
  \BibitemOpen
  \bibfield  {author} {\bibinfo {author} {\bibfnamefont {P.}~\bibnamefont
  {San-Jose}}\ and\ \bibinfo {author} {\bibfnamefont {E.}~\bibnamefont
  {Prada}},\ }\href {\doibase 10.1103/PhysRevB.88.121408} {\bibfield  {journal}
  {\bibinfo  {journal} {Phys. Rev. B}\ }\textbf {\bibinfo {volume} {88}},\
  \bibinfo {pages} {121408} (\bibinfo {year} {2013})}\BibitemShut {NoStop}%
\bibitem [{\citenamefont {Efimkin}\ and\ \citenamefont
  {MacDonald}(2018)}]{Efimkin2018}%
  \BibitemOpen
  \bibfield  {author} {\bibinfo {author} {\bibfnamefont {D.~K.}\ \bibnamefont
  {Efimkin}}\ and\ \bibinfo {author} {\bibfnamefont {A.~H.}\ \bibnamefont
  {MacDonald}},\ }\href {\doibase 10.1103/PhysRevB.98.035404} {\bibfield
  {journal} {\bibinfo  {journal} {Phys. Rev. B}\ }\textbf {\bibinfo {volume}
  {98}},\ \bibinfo {pages} {035404} (\bibinfo {year} {2018})}\BibitemShut
  {NoStop}%
\bibitem [{\citenamefont {Pal}\ \emph {et~al.}(2019)\citenamefont {Pal},
  \citenamefont {Spitz},\ and\ \citenamefont {Kindermann}}]{Pal2018a}%
  \BibitemOpen
  \bibfield  {author} {\bibinfo {author} {\bibfnamefont {H.~K.}\ \bibnamefont
  {Pal}}, \bibinfo {author} {\bibfnamefont {S.}~\bibnamefont {Spitz}}, \ and\
  \bibinfo {author} {\bibfnamefont {M.}~\bibnamefont {Kindermann}},\ }\href
  {\doibase 10.1103/PhysRevLett.123.186402} {\bibfield  {journal} {\bibinfo
  {journal} {Phys. Rev. Lett.}\ }\textbf {\bibinfo {volume} {123}},\ \bibinfo
  {pages} {186402} (\bibinfo {year} {2019})}\BibitemShut {NoStop}%
\bibitem [{\citenamefont {Alden}\ \emph {et~al.}(2013)\citenamefont {Alden},
  \citenamefont {Tsen}, \citenamefont {Huang}, \citenamefont {Hovden},
  \citenamefont {Brown}, \citenamefont {Park}, \citenamefont {Muller},\ and\
  \citenamefont {McEuen}}]{Alden2013}%
  \BibitemOpen
  \bibfield  {author} {\bibinfo {author} {\bibfnamefont {J.~S.}\ \bibnamefont
  {Alden}}, \bibinfo {author} {\bibfnamefont {A.~W.}\ \bibnamefont {Tsen}},
  \bibinfo {author} {\bibfnamefont {P.~Y.}\ \bibnamefont {Huang}}, \bibinfo
  {author} {\bibfnamefont {R.}~\bibnamefont {Hovden}}, \bibinfo {author}
  {\bibfnamefont {L.}~\bibnamefont {Brown}}, \bibinfo {author} {\bibfnamefont
  {J.}~\bibnamefont {Park}}, \bibinfo {author} {\bibfnamefont {D.~A.}\
  \bibnamefont {Muller}}, \ and\ \bibinfo {author} {\bibfnamefont {P.~L.}\
  \bibnamefont {McEuen}},\ }\href {\doibase 10.1073/pnas.1309394110} {\bibfield
   {journal} {\bibinfo  {journal} {Proc. Natl. Acad. Sci.}\ }\textbf {\bibinfo
  {volume} {110}},\ \bibinfo {pages} {11256} (\bibinfo {year}
  {2013})}\BibitemShut {NoStop}%
\bibitem [{\citenamefont {Nam}\ and\ \citenamefont {Koshino}(2017)}]{Nam2017}%
  \BibitemOpen
  \bibfield  {author} {\bibinfo {author} {\bibfnamefont {N.~N.~T.}\
  \bibnamefont {Nam}}\ and\ \bibinfo {author} {\bibfnamefont {M.}~\bibnamefont
  {Koshino}},\ }\href {\doibase 10.1103/PhysRevB.96.075311} {\bibfield
  {journal} {\bibinfo  {journal} {Phys. Rev. B}\ }\textbf {\bibinfo {volume}
  {96}},\ \bibinfo {pages} {075311} (\bibinfo {year} {2017})}\BibitemShut
  {NoStop}%
\bibitem [{\citenamefont {Andelkovi\'{c}}\ \emph {et~al.}(2018)\citenamefont
  {Andelkovi\'{c}}, \citenamefont {Covaci},\ and\ \citenamefont
  {Peeters}}]{Anelkovic2017}%
  \BibitemOpen
  \bibfield  {author} {\bibinfo {author} {\bibfnamefont {M.}~\bibnamefont
  {Andelkovi\'{c}}}, \bibinfo {author} {\bibfnamefont {L.}~\bibnamefont
  {Covaci}}, \ and\ \bibinfo {author} {\bibfnamefont {F.~M.}\ \bibnamefont
  {Peeters}},\ }\href {\doibase 10.1103/PhysRevMaterials.2.034004} {\bibfield
  {journal} {\bibinfo  {journal} {Phys. Rev. Mater.}\ }\textbf {\bibinfo
  {volume} {2}},\ \bibinfo {pages} {034004} (\bibinfo {year}
  {2018})}\BibitemShut {NoStop}%
\bibitem [{\citenamefont {Gargiulo}\ and\ \citenamefont
  {Yazyev}(2017)}]{Gargiulo2018}%
  \BibitemOpen
  \bibfield  {author} {\bibinfo {author} {\bibfnamefont {F.}~\bibnamefont
  {Gargiulo}}\ and\ \bibinfo {author} {\bibfnamefont {O.~V.}\ \bibnamefont
  {Yazyev}},\ }\href {\doibase 10.1088/2053-1583/aa9640} {\bibfield  {journal}
  {\bibinfo  {journal} {2D Mater.}\ }\textbf {\bibinfo {volume} {5}},\ \bibinfo
  {pages} {015019} (\bibinfo {year} {2017})}\BibitemShut {NoStop}%
\bibitem [{\citenamefont {Lucignano}\ \emph {et~al.}(2019)\citenamefont
  {Lucignano}, \citenamefont {Alf{\`{e}}}, \citenamefont {Cataudella},
  \citenamefont {Ninno},\ and\ \citenamefont {Cantele}}]{Lucignano2019}%
  \BibitemOpen
  \bibfield  {author} {\bibinfo {author} {\bibfnamefont {P.}~\bibnamefont
  {Lucignano}}, \bibinfo {author} {\bibfnamefont {D.}~\bibnamefont
  {Alf{\`{e}}}}, \bibinfo {author} {\bibfnamefont {V.}~\bibnamefont
  {Cataudella}}, \bibinfo {author} {\bibfnamefont {D.}~\bibnamefont {Ninno}}, \
  and\ \bibinfo {author} {\bibfnamefont {G.}~\bibnamefont {Cantele}},\ }\href
  {\doibase 10.1103/PhysRevB.99.195419} {\bibfield  {journal} {\bibinfo
  {journal} {Phys. Rev. B}\ }\textbf {\bibinfo {volume} {99}},\ \bibinfo
  {pages} {195419} (\bibinfo {year} {2019})}\BibitemShut {NoStop}%
\bibitem [{\citenamefont {Walet}\ and\ \citenamefont
  {Guinea}(2019)}]{Walet2019a}%
  \BibitemOpen
  \bibfield  {author} {\bibinfo {author} {\bibfnamefont {N.~R.}\ \bibnamefont
  {Walet}}\ and\ \bibinfo {author} {\bibfnamefont {F.}~\bibnamefont {Guinea}},\
  }\href {\doibase 10.1088/2053-1583/ab57f8} {\bibfield  {journal} {\bibinfo
  {journal} {2D Mater.}\ }\textbf {\bibinfo {volume} {7}},\ \bibinfo {pages}
  {015023} (\bibinfo {year} {2019})}\BibitemShut {NoStop}%
\bibitem [{\citenamefont {Ju}\ \emph {et~al.}(2015)\citenamefont {Ju},
  \citenamefont {Shi}, \citenamefont {Nair}, \citenamefont {Lv}, \citenamefont
  {Jin}, \citenamefont {Velasco}, \citenamefont {Ojeda-Aristizabal},
  \citenamefont {Bechtel}, \citenamefont {Martin}, \citenamefont {Zettl},
  \citenamefont {Analytis},\ and\ \citenamefont {Wang}}]{Ju2015}%
  \BibitemOpen
  \bibfield  {author} {\bibinfo {author} {\bibfnamefont {L.}~\bibnamefont
  {Ju}}, \bibinfo {author} {\bibfnamefont {Z.}~\bibnamefont {Shi}}, \bibinfo
  {author} {\bibfnamefont {N.}~\bibnamefont {Nair}}, \bibinfo {author}
  {\bibfnamefont {Y.}~\bibnamefont {Lv}}, \bibinfo {author} {\bibfnamefont
  {C.}~\bibnamefont {Jin}}, \bibinfo {author} {\bibfnamefont {J.}~\bibnamefont
  {Velasco}}, \bibinfo {author} {\bibfnamefont {C.}~\bibnamefont
  {Ojeda-Aristizabal}}, \bibinfo {author} {\bibfnamefont {H.~A.}\ \bibnamefont
  {Bechtel}}, \bibinfo {author} {\bibfnamefont {M.~C.}\ \bibnamefont {Martin}},
  \bibinfo {author} {\bibfnamefont {A.}~\bibnamefont {Zettl}}, \bibinfo
  {author} {\bibfnamefont {J.}~\bibnamefont {Analytis}}, \ and\ \bibinfo
  {author} {\bibfnamefont {F.}~\bibnamefont {Wang}},\ }\href {\doibase
  10.1038/nature14364} {\bibfield  {journal} {\bibinfo  {journal} {Nature}\
  }\textbf {\bibinfo {volume} {520}},\ \bibinfo {pages} {650} (\bibinfo {year}
  {2015})}\BibitemShut {NoStop}%
\bibitem [{\citenamefont {Yin}\ \emph {et~al.}(2016)\citenamefont {Yin},
  \citenamefont {Jiang}, \citenamefont {Qiao},\ and\ \citenamefont
  {He}}]{Yin2016}%
  \BibitemOpen
  \bibfield  {author} {\bibinfo {author} {\bibfnamefont {L.-J.}\ \bibnamefont
  {Yin}}, \bibinfo {author} {\bibfnamefont {H.}~\bibnamefont {Jiang}}, \bibinfo
  {author} {\bibfnamefont {J.-B.}\ \bibnamefont {Qiao}}, \ and\ \bibinfo
  {author} {\bibfnamefont {L.}~\bibnamefont {He}},\ }\href {\doibase
  10.1038/ncomms11760} {\bibfield  {journal} {\bibinfo  {journal} {Nat.
  Commun.}\ }\textbf {\bibinfo {volume} {7}},\ \bibinfo {pages} {11760}
  (\bibinfo {year} {2016})}\BibitemShut {NoStop}%
\bibitem [{\citenamefont {Huang}\ \emph {et~al.}(2018)\citenamefont {Huang},
  \citenamefont {Kim}, \citenamefont {Efimkin}, \citenamefont {Lovorn},
  \citenamefont {Taniguchi}, \citenamefont {Watanabe}, \citenamefont
  {MacDonald}, \citenamefont {Tutuc},\ and\ \citenamefont {LeRoy}}]{Huang2018}%
  \BibitemOpen
  \bibfield  {author} {\bibinfo {author} {\bibfnamefont {S.}~\bibnamefont
  {Huang}}, \bibinfo {author} {\bibfnamefont {K.}~\bibnamefont {Kim}}, \bibinfo
  {author} {\bibfnamefont {D.~K.}\ \bibnamefont {Efimkin}}, \bibinfo {author}
  {\bibfnamefont {T.}~\bibnamefont {Lovorn}}, \bibinfo {author} {\bibfnamefont
  {T.}~\bibnamefont {Taniguchi}}, \bibinfo {author} {\bibfnamefont
  {K.}~\bibnamefont {Watanabe}}, \bibinfo {author} {\bibfnamefont {A.~H.}\
  \bibnamefont {MacDonald}}, \bibinfo {author} {\bibfnamefont {E.}~\bibnamefont
  {Tutuc}}, \ and\ \bibinfo {author} {\bibfnamefont {B.~J.}\ \bibnamefont
  {LeRoy}},\ }\href {\doibase 10.1103/PhysRevLett.121.037702} {\bibfield
  {journal} {\bibinfo  {journal} {Phys. Rev. Lett.}\ }\textbf {\bibinfo
  {volume} {121}},\ \bibinfo {pages} {037702} (\bibinfo {year}
  {2018})}\BibitemShut {NoStop}%
\bibitem [{\citenamefont {Sunku}\ \emph {et~al.}(2018)\citenamefont {Sunku},
  \citenamefont {Ni}, \citenamefont {Jiang}, \citenamefont {Yoo}, \citenamefont
  {Sternbach}, \citenamefont {McLeod}, \citenamefont {Stauber}, \citenamefont
  {Xiong}, \citenamefont {Taniguchi}, \citenamefont {Watanabe}, \citenamefont
  {Kim}, \citenamefont {Fogler},\ and\ \citenamefont {Basov}}]{Sunku2018}%
  \BibitemOpen
  \bibfield  {author} {\bibinfo {author} {\bibfnamefont {S.~S.}\ \bibnamefont
  {Sunku}}, \bibinfo {author} {\bibfnamefont {G.~X.}\ \bibnamefont {Ni}},
  \bibinfo {author} {\bibfnamefont {B.~Y.}\ \bibnamefont {Jiang}}, \bibinfo
  {author} {\bibfnamefont {H.}~\bibnamefont {Yoo}}, \bibinfo {author}
  {\bibfnamefont {A.}~\bibnamefont {Sternbach}}, \bibinfo {author}
  {\bibfnamefont {A.~S.}\ \bibnamefont {McLeod}}, \bibinfo {author}
  {\bibfnamefont {T.}~\bibnamefont {Stauber}}, \bibinfo {author} {\bibfnamefont
  {L.}~\bibnamefont {Xiong}}, \bibinfo {author} {\bibfnamefont
  {T.}~\bibnamefont {Taniguchi}}, \bibinfo {author} {\bibfnamefont
  {K.}~\bibnamefont {Watanabe}}, \bibinfo {author} {\bibfnamefont
  {P.}~\bibnamefont {Kim}}, \bibinfo {author} {\bibfnamefont {M.~M.}\
  \bibnamefont {Fogler}}, \ and\ \bibinfo {author} {\bibfnamefont {D.~N.}\
  \bibnamefont {Basov}},\ }\href {\doibase 10.1126/science.aau5144} {\bibfield
  {journal} {\bibinfo  {journal} {Science}\ }\textbf {\bibinfo {volume}
  {362}},\ \bibinfo {pages} {1153} (\bibinfo {year} {2018})}\BibitemShut
  {NoStop}%
\bibitem [{\citenamefont {Rickhaus}\ \emph {et~al.}(2018)\citenamefont
  {Rickhaus}, \citenamefont {Wallbank}, \citenamefont {Slizovskiy},
  \citenamefont {Pisoni}, \citenamefont {Overweg}, \citenamefont {Lee},
  \citenamefont {Eich}, \citenamefont {Liu}, \citenamefont {Watanabe},
  \citenamefont {Taniguchi}, \citenamefont {Ihn},\ and\ \citenamefont
  {Ensslin}}]{Rickhaus2018}%
  \BibitemOpen
  \bibfield  {author} {\bibinfo {author} {\bibfnamefont {P.}~\bibnamefont
  {Rickhaus}}, \bibinfo {author} {\bibfnamefont {J.}~\bibnamefont {Wallbank}},
  \bibinfo {author} {\bibfnamefont {S.}~\bibnamefont {Slizovskiy}}, \bibinfo
  {author} {\bibfnamefont {R.}~\bibnamefont {Pisoni}}, \bibinfo {author}
  {\bibfnamefont {H.}~\bibnamefont {Overweg}}, \bibinfo {author} {\bibfnamefont
  {Y.}~\bibnamefont {Lee}}, \bibinfo {author} {\bibfnamefont {M.}~\bibnamefont
  {Eich}}, \bibinfo {author} {\bibfnamefont {M.-H.}\ \bibnamefont {Liu}},
  \bibinfo {author} {\bibfnamefont {K.}~\bibnamefont {Watanabe}}, \bibinfo
  {author} {\bibfnamefont {T.}~\bibnamefont {Taniguchi}}, \bibinfo {author}
  {\bibfnamefont {T.}~\bibnamefont {Ihn}}, \ and\ \bibinfo {author}
  {\bibfnamefont {K.}~\bibnamefont {Ensslin}},\ }\href {\doibase
  10.1021/acs.nanolett.8b02387} {\bibfield  {journal} {\bibinfo  {journal}
  {Nano Lett.}\ }\textbf {\bibinfo {volume} {18}},\ \bibinfo {pages} {6725}
  (\bibinfo {year} {2018})}\BibitemShut {NoStop}%
\bibitem [{\citenamefont {Yoo}\ \emph {et~al.}(2019)\citenamefont {Yoo},
  \citenamefont {Engelke}, \citenamefont {Carr}, \citenamefont {Fang},
  \citenamefont {Zhang}, \citenamefont {Cazeaux}, \citenamefont {Sung},
  \citenamefont {Hovden}, \citenamefont {Tsen}, \citenamefont {Taniguchi},
  \citenamefont {Watanabe}, \citenamefont {Yi}, \citenamefont {Kim},
  \citenamefont {Luskin}, \citenamefont {Tadmor}, \citenamefont {Kaxiras},\
  and\ \citenamefont {Kim}}]{Yoo2018}%
  \BibitemOpen
  \bibfield  {author} {\bibinfo {author} {\bibfnamefont {H.}~\bibnamefont
  {Yoo}}, \bibinfo {author} {\bibfnamefont {R.}~\bibnamefont {Engelke}},
  \bibinfo {author} {\bibfnamefont {S.}~\bibnamefont {Carr}}, \bibinfo {author}
  {\bibfnamefont {S.}~\bibnamefont {Fang}}, \bibinfo {author} {\bibfnamefont
  {K.}~\bibnamefont {Zhang}}, \bibinfo {author} {\bibfnamefont
  {P.}~\bibnamefont {Cazeaux}}, \bibinfo {author} {\bibfnamefont {S.~H.}\
  \bibnamefont {Sung}}, \bibinfo {author} {\bibfnamefont {R.}~\bibnamefont
  {Hovden}}, \bibinfo {author} {\bibfnamefont {A.~W.}\ \bibnamefont {Tsen}},
  \bibinfo {author} {\bibfnamefont {T.}~\bibnamefont {Taniguchi}}, \bibinfo
  {author} {\bibfnamefont {K.}~\bibnamefont {Watanabe}}, \bibinfo {author}
  {\bibfnamefont {G.-C.}\ \bibnamefont {Yi}}, \bibinfo {author} {\bibfnamefont
  {M.}~\bibnamefont {Kim}}, \bibinfo {author} {\bibfnamefont {M.}~\bibnamefont
  {Luskin}}, \bibinfo {author} {\bibfnamefont {E.~B.}\ \bibnamefont {Tadmor}},
  \bibinfo {author} {\bibfnamefont {E.}~\bibnamefont {Kaxiras}}, \ and\
  \bibinfo {author} {\bibfnamefont {P.}~\bibnamefont {Kim}},\ }\href {\doibase
  10.1038/s41563-019-0346-z} {\bibfield  {journal} {\bibinfo  {journal} {Nat.
  Mater.}\ }\textbf {\bibinfo {volume} {18}},\ \bibinfo {pages} {448} (\bibinfo
  {year} {2019})}\BibitemShut {NoStop}%
\bibitem [{\citenamefont {Xu}\ \emph {et~al.}(2019)\citenamefont {Xu},
  \citenamefont {Berdyugin}, \citenamefont {Kumaravadivel}, \citenamefont
  {Guinea}, \citenamefont {Kumar}, \citenamefont {Bandurin}, \citenamefont
  {Morozov}, \citenamefont {Kuang}, \citenamefont {Tsim}, \citenamefont {Liu},
  \citenamefont {Edgar}, \citenamefont {Grigorieva}, \citenamefont {Fal'ko},
  \citenamefont {Kim},\ and\ \citenamefont {Geim}}]{Xu2019}%
  \BibitemOpen
  \bibfield  {author} {\bibinfo {author} {\bibfnamefont {S.~G.}\ \bibnamefont
  {Xu}}, \bibinfo {author} {\bibfnamefont {A.~I.}\ \bibnamefont {Berdyugin}},
  \bibinfo {author} {\bibfnamefont {P.}~\bibnamefont {Kumaravadivel}}, \bibinfo
  {author} {\bibfnamefont {F.}~\bibnamefont {Guinea}}, \bibinfo {author}
  {\bibfnamefont {R.~K.}\ \bibnamefont {Kumar}}, \bibinfo {author}
  {\bibfnamefont {D.~A.}\ \bibnamefont {Bandurin}}, \bibinfo {author}
  {\bibfnamefont {S.~V.}\ \bibnamefont {Morozov}}, \bibinfo {author}
  {\bibfnamefont {W.}~\bibnamefont {Kuang}}, \bibinfo {author} {\bibfnamefont
  {B.}~\bibnamefont {Tsim}}, \bibinfo {author} {\bibfnamefont {S.}~\bibnamefont
  {Liu}}, \bibinfo {author} {\bibfnamefont {J.~H.}\ \bibnamefont {Edgar}},
  \bibinfo {author} {\bibfnamefont {I.~V.}\ \bibnamefont {Grigorieva}},
  \bibinfo {author} {\bibfnamefont {V.~I.}\ \bibnamefont {Fal'ko}}, \bibinfo
  {author} {\bibfnamefont {M.}~\bibnamefont {Kim}}, \ and\ \bibinfo {author}
  {\bibfnamefont {A.~K.}\ \bibnamefont {Geim}},\ }\href {\doibase
  10.1038/s41467-019-11971-7} {\bibfield  {journal} {\bibinfo  {journal} {Nat.
  Commun.}\ }\textbf {\bibinfo {volume} {10}},\ \bibinfo {pages} {4008}
  (\bibinfo {year} {2019})}\BibitemShut {NoStop}%
\bibitem [{\citenamefont {Yankowitz}\ \emph {et~al.}(2019)\citenamefont
  {Yankowitz}, \citenamefont {Chen}, \citenamefont {Polshyn}, \citenamefont
  {Zhang}, \citenamefont {Watanabe}, \citenamefont {Taniguchi}, \citenamefont
  {Graf}, \citenamefont {Young},\ and\ \citenamefont {Dean}}]{Yankowitz2019}%
  \BibitemOpen
  \bibfield  {author} {\bibinfo {author} {\bibfnamefont {M.}~\bibnamefont
  {Yankowitz}}, \bibinfo {author} {\bibfnamefont {S.}~\bibnamefont {Chen}},
  \bibinfo {author} {\bibfnamefont {H.}~\bibnamefont {Polshyn}}, \bibinfo
  {author} {\bibfnamefont {Y.}~\bibnamefont {Zhang}}, \bibinfo {author}
  {\bibfnamefont {K.}~\bibnamefont {Watanabe}}, \bibinfo {author}
  {\bibfnamefont {T.}~\bibnamefont {Taniguchi}}, \bibinfo {author}
  {\bibfnamefont {D.}~\bibnamefont {Graf}}, \bibinfo {author} {\bibfnamefont
  {A.~F.}\ \bibnamefont {Young}}, \ and\ \bibinfo {author} {\bibfnamefont
  {C.~R.}\ \bibnamefont {Dean}},\ }\href {\doibase 10.1126/science.aav1910}
  {\bibfield  {journal} {\bibinfo  {journal} {Science}\ }\textbf {\bibinfo
  {volume} {363}},\ \bibinfo {pages} {1059} (\bibinfo {year}
  {2019})}\BibitemShut {NoStop}%
\bibitem [{\citenamefont {Wu}\ \emph {et~al.}(2019)\citenamefont {Wu},
  \citenamefont {Jian},\ and\ \citenamefont {Xu}}]{Wu2019}%
  \BibitemOpen
  \bibfield  {author} {\bibinfo {author} {\bibfnamefont {X.-C.}\ \bibnamefont
  {Wu}}, \bibinfo {author} {\bibfnamefont {C.-M.}\ \bibnamefont {Jian}}, \ and\
  \bibinfo {author} {\bibfnamefont {C.}~\bibnamefont {Xu}},\ }\href {\doibase
  10.1103/PhysRevB.99.161405} {\bibfield  {journal} {\bibinfo  {journal} {Phys.
  Rev. B}\ }\textbf {\bibinfo {volume} {99}},\ \bibinfo {pages} {161405}
  (\bibinfo {year} {2019})}\BibitemShut {NoStop}%
\bibitem [{\citenamefont {Giamarchi}(1997)}]{Giamarchi1997}%
  \BibitemOpen
  \bibfield  {author} {\bibinfo {author} {\bibfnamefont {T.}~\bibnamefont
  {Giamarchi}},\ }\href {\doibase
  https://doi.org/10.1016/S0921-4526(96)00768-5} {\bibfield  {journal}
  {\bibinfo  {journal} {Physica B: Condensed Matter}\ }\textbf {\bibinfo
  {volume} {230-232}},\ \bibinfo {pages} {975 } (\bibinfo {year} {1997})},\
  \bibinfo {note} {proceedings of the International Conference on Strongly
  Correlated Electron Systems}\BibitemShut {NoStop}%
\bibitem [{\citenamefont {Emery}(1979)}]{Emery1979}%
  \BibitemOpen
  \bibfield  {author} {\bibinfo {author} {\bibfnamefont {V.~J.}\ \bibnamefont
  {Emery}},\ }\enquote {\bibinfo {title} {Theory of the one-dimensional
  electron gas},}\ in\ \href {\doibase 10.1007/978-1-4613-2895-7_6} {\emph
  {\bibinfo {booktitle} {Highly Conducting One-Dimensional Solids}}},\ \bibinfo
  {editor} {edited by\ \bibinfo {editor} {\bibfnamefont {J.~T.}\ \bibnamefont
  {Devreese}}, \bibinfo {editor} {\bibfnamefont {R.~P.}\ \bibnamefont
  {Evrard}}, \ and\ \bibinfo {editor} {\bibfnamefont {V.~E.}\ \bibnamefont {van
  Doren}}}\ (\bibinfo  {publisher} {Springer US},\ \bibinfo {address} {Boston,
  MA},\ \bibinfo {year} {1979})\ pp.\ \bibinfo {pages} {247--303}\BibitemShut
  {NoStop}%
\bibitem [{\citenamefont {{Cao}}\ \emph {et~al.}(2019)\citenamefont {{Cao}},
  \citenamefont {{Chowdhury}}, \citenamefont {{Rodan-Legrain}}, \citenamefont
  {{Rubies-Bigord{\`a}}}, \citenamefont {{Watanabe}}, \citenamefont
  {{Taniguchi}}, \citenamefont {{Senthil}},\ and\ \citenamefont
  {{Jarillo-Herrero}}}]{Cao2019}%
  \BibitemOpen
  \bibfield  {author} {\bibinfo {author} {\bibfnamefont {Y.}~\bibnamefont
  {{Cao}}}, \bibinfo {author} {\bibfnamefont {D.}~\bibnamefont {{Chowdhury}}},
  \bibinfo {author} {\bibfnamefont {D.}~\bibnamefont {{Rodan-Legrain}}},
  \bibinfo {author} {\bibfnamefont {O.}~\bibnamefont {{Rubies-Bigord{\`a}}}},
  \bibinfo {author} {\bibfnamefont {K.}~\bibnamefont {{Watanabe}}}, \bibinfo
  {author} {\bibfnamefont {T.}~\bibnamefont {{Taniguchi}}}, \bibinfo {author}
  {\bibfnamefont {T.}~\bibnamefont {{Senthil}}}, \ and\ \bibinfo {author}
  {\bibfnamefont {P.}~\bibnamefont {{Jarillo-Herrero}}},\ }\href@noop {}
  {\bibfield  {journal} {\bibinfo  {journal} {arXiv preprint}\ } (\bibinfo
  {year} {2019})},\ \Eprint {http://arxiv.org/abs/1901.03710} {1901.03710}
  \BibitemShut {NoStop}%
\bibitem [{\citenamefont {Jiang}\ \emph {et~al.}(2019)\citenamefont {Jiang},
  \citenamefont {Lai}, \citenamefont {Watanabe}, \citenamefont {Taniguchi},
  \citenamefont {Haule}, \citenamefont {Mao},\ and\ \citenamefont
  {Andrei}}]{Jiang2019}%
  \BibitemOpen
  \bibfield  {author} {\bibinfo {author} {\bibfnamefont {Y.}~\bibnamefont
  {Jiang}}, \bibinfo {author} {\bibfnamefont {X.}~\bibnamefont {Lai}}, \bibinfo
  {author} {\bibfnamefont {K.}~\bibnamefont {Watanabe}}, \bibinfo {author}
  {\bibfnamefont {T.}~\bibnamefont {Taniguchi}}, \bibinfo {author}
  {\bibfnamefont {K.}~\bibnamefont {Haule}}, \bibinfo {author} {\bibfnamefont
  {J.}~\bibnamefont {Mao}}, \ and\ \bibinfo {author} {\bibfnamefont {E.~Y.}\
  \bibnamefont {Andrei}},\ }\href {\doibase 10.1038/s41586-019-1460-4}
  {\bibfield  {journal} {\bibinfo  {journal} {Nature}\ } (\bibinfo {year}
  {2019}),\ 10.1038/s41586-019-1460-4},\ \Eprint
  {http://arxiv.org/abs/1904.10153} {1904.10153} \BibitemShut {NoStop}%
\bibitem [{\citenamefont {Kerelsky}\ \emph {et~al.}(2019)\citenamefont
  {Kerelsky}, \citenamefont {McGilly}, \citenamefont {Kennes}, \citenamefont
  {Xian}, \citenamefont {Yankowitz}, \citenamefont {Chen}, \citenamefont
  {Watanabe}, \citenamefont {Taniguchi}, \citenamefont {Hone}, \citenamefont
  {Dean}, \citenamefont {Rubio},\ and\ \citenamefont
  {Pasupathy}}]{Kerelsky2019}%
  \BibitemOpen
  \bibfield  {author} {\bibinfo {author} {\bibfnamefont {A.}~\bibnamefont
  {Kerelsky}}, \bibinfo {author} {\bibfnamefont {L.~J.}\ \bibnamefont
  {McGilly}}, \bibinfo {author} {\bibfnamefont {D.~M.}\ \bibnamefont {Kennes}},
  \bibinfo {author} {\bibfnamefont {L.}~\bibnamefont {Xian}}, \bibinfo {author}
  {\bibfnamefont {M.}~\bibnamefont {Yankowitz}}, \bibinfo {author}
  {\bibfnamefont {S.}~\bibnamefont {Chen}}, \bibinfo {author} {\bibfnamefont
  {K.}~\bibnamefont {Watanabe}}, \bibinfo {author} {\bibfnamefont
  {T.}~\bibnamefont {Taniguchi}}, \bibinfo {author} {\bibfnamefont
  {J.}~\bibnamefont {Hone}}, \bibinfo {author} {\bibfnamefont {C.}~\bibnamefont
  {Dean}}, \bibinfo {author} {\bibfnamefont {A.}~\bibnamefont {Rubio}}, \ and\
  \bibinfo {author} {\bibfnamefont {A.~N.}\ \bibnamefont {Pasupathy}},\ }\href
  {\doibase 10.1038/s41586-019-1431-9} {\bibfield  {journal} {\bibinfo
  {journal} {Nature}\ }\textbf {\bibinfo {volume} {572}},\ \bibinfo {pages}
  {95} (\bibinfo {year} {2019})}\BibitemShut {NoStop}%
\bibitem [{\citenamefont {Choi}\ \emph {et~al.}(2019)\citenamefont {Choi},
  \citenamefont {Kemmer}, \citenamefont {Peng}, \citenamefont {Thomson},
  \citenamefont {Arora}, \citenamefont {Polski}, \citenamefont {Zhang},
  \citenamefont {Ren}, \citenamefont {Alicea}, \citenamefont {Refael},
  \citenamefont {von Oppen}, \citenamefont {Watanabe}, \citenamefont
  {Taniguchi},\ and\ \citenamefont {Nadj-Perge}}]{Choi2019a}%
  \BibitemOpen
  \bibfield  {author} {\bibinfo {author} {\bibfnamefont {Y.}~\bibnamefont
  {Choi}}, \bibinfo {author} {\bibfnamefont {J.}~\bibnamefont {Kemmer}},
  \bibinfo {author} {\bibfnamefont {Y.}~\bibnamefont {Peng}}, \bibinfo {author}
  {\bibfnamefont {A.}~\bibnamefont {Thomson}}, \bibinfo {author} {\bibfnamefont
  {H.}~\bibnamefont {Arora}}, \bibinfo {author} {\bibfnamefont
  {R.}~\bibnamefont {Polski}}, \bibinfo {author} {\bibfnamefont
  {Y.}~\bibnamefont {Zhang}}, \bibinfo {author} {\bibfnamefont
  {H.}~\bibnamefont {Ren}}, \bibinfo {author} {\bibfnamefont {J.}~\bibnamefont
  {Alicea}}, \bibinfo {author} {\bibfnamefont {G.}~\bibnamefont {Refael}},
  \bibinfo {author} {\bibfnamefont {F.}~\bibnamefont {von Oppen}}, \bibinfo
  {author} {\bibfnamefont {K.}~\bibnamefont {Watanabe}}, \bibinfo {author}
  {\bibfnamefont {T.}~\bibnamefont {Taniguchi}}, \ and\ \bibinfo {author}
  {\bibfnamefont {S.}~\bibnamefont {Nadj-Perge}},\ }\href {\doibase
  10.1038/s41567-019-0606-5} {\bibfield  {journal} {\bibinfo  {journal} {Nat.
  Phys.}\ } (\bibinfo {year} {2019}),\ 10.1038/s41567-019-0606-5}\BibitemShut
  {NoStop}%
\bibitem [{\citenamefont {Xie}\ \emph {et~al.}(2019)\citenamefont {Xie},
  \citenamefont {Lian}, \citenamefont {J{\"{a}}ck}, \citenamefont {Liu},
  \citenamefont {Chiu}, \citenamefont {Watanabe}, \citenamefont {Taniguchi},
  \citenamefont {Bernevig},\ and\ \citenamefont {Yazdani}}]{Xie2019}%
  \BibitemOpen
  \bibfield  {author} {\bibinfo {author} {\bibfnamefont {Y.}~\bibnamefont
  {Xie}}, \bibinfo {author} {\bibfnamefont {B.}~\bibnamefont {Lian}}, \bibinfo
  {author} {\bibfnamefont {B.}~\bibnamefont {J{\"{a}}ck}}, \bibinfo {author}
  {\bibfnamefont {X.}~\bibnamefont {Liu}}, \bibinfo {author} {\bibfnamefont
  {C.-l.}\ \bibnamefont {Chiu}}, \bibinfo {author} {\bibfnamefont
  {K.}~\bibnamefont {Watanabe}}, \bibinfo {author} {\bibfnamefont
  {T.}~\bibnamefont {Taniguchi}}, \bibinfo {author} {\bibfnamefont {B.~A.}\
  \bibnamefont {Bernevig}}, \ and\ \bibinfo {author} {\bibfnamefont
  {A.}~\bibnamefont {Yazdani}},\ }\href {\doibase 10.1038/s41586-019-1422-x}
  {\bibfield  {journal} {\bibinfo  {journal} {Nature}\ }\textbf {\bibinfo
  {volume} {572}},\ \bibinfo {pages} {101} (\bibinfo {year}
  {2019})}\BibitemShut {NoStop}%
\bibitem [{\citenamefont {van Wijk}\ \emph {et~al.}(2015)\citenamefont {van
  Wijk}, \citenamefont {Schuring}, \citenamefont {Katsnelson},\ and\
  \citenamefont {Fasolino}}]{Wijk2015}%
  \BibitemOpen
  \bibfield  {author} {\bibinfo {author} {\bibfnamefont {M.~M.}\ \bibnamefont
  {van Wijk}}, \bibinfo {author} {\bibfnamefont {A.}~\bibnamefont {Schuring}},
  \bibinfo {author} {\bibfnamefont {M.~I.}\ \bibnamefont {Katsnelson}}, \ and\
  \bibinfo {author} {\bibfnamefont {A.}~\bibnamefont {Fasolino}},\ }\href
  {\doibase 10.1088/2053-1583/2/3/034010} {\bibfield  {journal} {\bibinfo
  {journal} {2D Mater.}\ }\textbf {\bibinfo {volume} {2}},\ \bibinfo {pages}
  {034010} (\bibinfo {year} {2015})}\BibitemShut {NoStop}%
\bibitem [{\citenamefont {Zhang}\ and\ \citenamefont
  {Tadmor}(2018)}]{Zhang2018b}%
  \BibitemOpen
  \bibfield  {author} {\bibinfo {author} {\bibfnamefont {K.}~\bibnamefont
  {Zhang}}\ and\ \bibinfo {author} {\bibfnamefont {E.~B.}\ \bibnamefont
  {Tadmor}},\ }\href {\doibase 10.1016/j.jmps.2017.12.005} {\bibfield
  {journal} {\bibinfo  {journal} {J. Mech. Phys. Solids}\ }\textbf {\bibinfo
  {volume} {112}},\ \bibinfo {pages} {225} (\bibinfo {year}
  {2018})}\BibitemShut {NoStop}%
\bibitem [{\citenamefont {Li}\ \emph {et~al.}(2016)\citenamefont {Li},
  \citenamefont {Wang}, \citenamefont {McFaul}, \citenamefont {Zern},
  \citenamefont {Ren}, \citenamefont {Watanabe}, \citenamefont {Taniguchi},
  \citenamefont {Qiao},\ and\ \citenamefont {Zhu}}]{Li2016d}%
  \BibitemOpen
  \bibfield  {author} {\bibinfo {author} {\bibfnamefont {J.}~\bibnamefont
  {Li}}, \bibinfo {author} {\bibfnamefont {K.}~\bibnamefont {Wang}}, \bibinfo
  {author} {\bibfnamefont {K.~J.}\ \bibnamefont {McFaul}}, \bibinfo {author}
  {\bibfnamefont {Z.}~\bibnamefont {Zern}}, \bibinfo {author} {\bibfnamefont
  {Y.}~\bibnamefont {Ren}}, \bibinfo {author} {\bibfnamefont {K.}~\bibnamefont
  {Watanabe}}, \bibinfo {author} {\bibfnamefont {T.}~\bibnamefont {Taniguchi}},
  \bibinfo {author} {\bibfnamefont {Z.}~\bibnamefont {Qiao}}, \ and\ \bibinfo
  {author} {\bibfnamefont {J.}~\bibnamefont {Zhu}},\ }\href {\doibase
  10.1038/nnano.2016.158} {\bibfield  {journal} {\bibinfo  {journal} {Nat.
  Nanotechnol.}\ }\textbf {\bibinfo {volume} {11}},\ \bibinfo {pages} {1060}
  (\bibinfo {year} {2016})}\BibitemShut {NoStop}%
\bibitem [{\citenamefont {Hunt}\ \emph {et~al.}(2013)\citenamefont {Hunt},
  \citenamefont {Sanchez-Yamagishi}, \citenamefont {Young}, \citenamefont
  {Yankowitz}, \citenamefont {LeRoy}, \citenamefont {Watanabe}, \citenamefont
  {Taniguchi}, \citenamefont {Moon}, \citenamefont {Koshino}, \citenamefont
  {Jarillo-Herrero},\ and\ \citenamefont {Ashoori}}]{Hunt2013}%
  \BibitemOpen
  \bibfield  {author} {\bibinfo {author} {\bibfnamefont {B.}~\bibnamefont
  {Hunt}}, \bibinfo {author} {\bibfnamefont {J.~D.}\ \bibnamefont
  {Sanchez-Yamagishi}}, \bibinfo {author} {\bibfnamefont {A.~F.}\ \bibnamefont
  {Young}}, \bibinfo {author} {\bibfnamefont {M.}~\bibnamefont {Yankowitz}},
  \bibinfo {author} {\bibfnamefont {B.~J.}\ \bibnamefont {LeRoy}}, \bibinfo
  {author} {\bibfnamefont {K.}~\bibnamefont {Watanabe}}, \bibinfo {author}
  {\bibfnamefont {T.}~\bibnamefont {Taniguchi}}, \bibinfo {author}
  {\bibfnamefont {P.}~\bibnamefont {Moon}}, \bibinfo {author} {\bibfnamefont
  {M.}~\bibnamefont {Koshino}}, \bibinfo {author} {\bibfnamefont
  {P.}~\bibnamefont {Jarillo-Herrero}}, \ and\ \bibinfo {author} {\bibfnamefont
  {R.~C.}\ \bibnamefont {Ashoori}},\ }\href {\doibase 10.1126/science.1237240}
  {\bibfield  {journal} {\bibinfo  {journal} {Science}\ }\textbf {\bibinfo
  {volume} {340}},\ \bibinfo {pages} {1427} (\bibinfo {year}
  {2013})}\BibitemShut {NoStop}%
\bibitem [{\citenamefont {Chen}\ \emph {et~al.}(2014)\citenamefont {Chen},
  \citenamefont {Shi}, \citenamefont {Yang}, \citenamefont {Lu}, \citenamefont
  {Lai}, \citenamefont {Yan}, \citenamefont {Wang}, \citenamefont {Zhang},\
  and\ \citenamefont {Li}}]{Chen2014b}%
  \BibitemOpen
  \bibfield  {author} {\bibinfo {author} {\bibfnamefont {Z.-G.}\ \bibnamefont
  {Chen}}, \bibinfo {author} {\bibfnamefont {Z.}~\bibnamefont {Shi}}, \bibinfo
  {author} {\bibfnamefont {W.}~\bibnamefont {Yang}}, \bibinfo {author}
  {\bibfnamefont {X.}~\bibnamefont {Lu}}, \bibinfo {author} {\bibfnamefont
  {Y.}~\bibnamefont {Lai}}, \bibinfo {author} {\bibfnamefont {H.}~\bibnamefont
  {Yan}}, \bibinfo {author} {\bibfnamefont {F.}~\bibnamefont {Wang}}, \bibinfo
  {author} {\bibfnamefont {G.}~\bibnamefont {Zhang}}, \ and\ \bibinfo {author}
  {\bibfnamefont {Z.}~\bibnamefont {Li}},\ }\href {\doibase 10.1038/ncomms5461}
  {\bibfield  {journal} {\bibinfo  {journal} {Nat. Commun.}\ }\textbf {\bibinfo
  {volume} {5}},\ \bibinfo {pages} {4461} (\bibinfo {year} {2014})}\BibitemShut
  {NoStop}%
\bibitem [{\citenamefont {Moon}\ and\ \citenamefont
  {Koshino}(2014)}]{Moon2014a}%
  \BibitemOpen
  \bibfield  {author} {\bibinfo {author} {\bibfnamefont {P.}~\bibnamefont
  {Moon}}\ and\ \bibinfo {author} {\bibfnamefont {M.}~\bibnamefont {Koshino}},\
  }\href {\doibase 10.1103/PhysRevB.90.155406} {\bibfield  {journal} {\bibinfo
  {journal} {Phys. Rev. B}\ }\textbf {\bibinfo {volume} {90}},\ \bibinfo
  {pages} {155406} (\bibinfo {year} {2014})}\BibitemShut {NoStop}%
\bibitem [{\citenamefont {Jung}\ \emph {et~al.}(2015)\citenamefont {Jung},
  \citenamefont {DaSilva}, \citenamefont {MacDonald},\ and\ \citenamefont
  {Adam}}]{Jung2015}%
  \BibitemOpen
  \bibfield  {author} {\bibinfo {author} {\bibfnamefont {J.}~\bibnamefont
  {Jung}}, \bibinfo {author} {\bibfnamefont {A.~M.}\ \bibnamefont {DaSilva}},
  \bibinfo {author} {\bibfnamefont {A.~H.}\ \bibnamefont {MacDonald}}, \ and\
  \bibinfo {author} {\bibfnamefont {S.}~\bibnamefont {Adam}},\ }\href {\doibase
  10.1038/ncomms7308} {\bibfield  {journal} {\bibinfo  {journal} {Nat.
  Commun.}\ }\textbf {\bibinfo {volume} {6}},\ \bibinfo {pages} {6308}
  (\bibinfo {year} {2015})}\BibitemShut {NoStop}%
\bibitem [{\citenamefont {Mucha-Kruczynski}\ \emph {et~al.}(2013)\citenamefont
  {Mucha-Kruczynski}, \citenamefont {Wallbank},\ and\ \citenamefont
  {Fal'ko}}]{Mucha-Kruczynski2013}%
  \BibitemOpen
  \bibfield  {author} {\bibinfo {author} {\bibfnamefont {M.}~\bibnamefont
  {Mucha-Kruczynski}}, \bibinfo {author} {\bibfnamefont {J.~R.}\ \bibnamefont
  {Wallbank}}, \ and\ \bibinfo {author} {\bibfnamefont {V.~I.}\ \bibnamefont
  {Fal'ko}},\ }\href {\doibase 10.1103/PhysRevB.88.205418} {\bibfield
  {journal} {\bibinfo  {journal} {Phys. Rev. B}\ }\textbf {\bibinfo {volume}
  {88}} (\bibinfo {year} {2013}),\ 10.1103/PhysRevB.88.205418}\BibitemShut
  {NoStop}%
\bibitem [{\citenamefont {Dean}\ \emph {et~al.}(2013)\citenamefont {Dean},
  \citenamefont {Wang}, \citenamefont {Maher}, \citenamefont {Forsythe},
  \citenamefont {Ghahari}, \citenamefont {Gao}, \citenamefont {Katoch},
  \citenamefont {Ishigami}, \citenamefont {Moon}, \citenamefont {Koshino},
  \citenamefont {Taniguchi}, \citenamefont {Watanabe}, \citenamefont {Shepard},
  \citenamefont {Hone},\ and\ \citenamefont {Kim}}]{Dean2013}%
  \BibitemOpen
  \bibfield  {author} {\bibinfo {author} {\bibfnamefont {C.~R.}\ \bibnamefont
  {Dean}}, \bibinfo {author} {\bibfnamefont {L.}~\bibnamefont {Wang}}, \bibinfo
  {author} {\bibfnamefont {P.}~\bibnamefont {Maher}}, \bibinfo {author}
  {\bibfnamefont {C.}~\bibnamefont {Forsythe}}, \bibinfo {author}
  {\bibfnamefont {F.}~\bibnamefont {Ghahari}}, \bibinfo {author} {\bibfnamefont
  {Y.}~\bibnamefont {Gao}}, \bibinfo {author} {\bibfnamefont {J.}~\bibnamefont
  {Katoch}}, \bibinfo {author} {\bibfnamefont {M.}~\bibnamefont {Ishigami}},
  \bibinfo {author} {\bibfnamefont {P.}~\bibnamefont {Moon}}, \bibinfo {author}
  {\bibfnamefont {M.}~\bibnamefont {Koshino}}, \bibinfo {author} {\bibfnamefont
  {T.}~\bibnamefont {Taniguchi}}, \bibinfo {author} {\bibfnamefont
  {K.}~\bibnamefont {Watanabe}}, \bibinfo {author} {\bibfnamefont {K.~L.}\
  \bibnamefont {Shepard}}, \bibinfo {author} {\bibfnamefont {J.}~\bibnamefont
  {Hone}}, \ and\ \bibinfo {author} {\bibfnamefont {P.}~\bibnamefont {Kim}},\
  }\href {\doibase 10.1038/nature12186} {\bibfield  {journal} {\bibinfo
  {journal} {Nature}\ }\textbf {\bibinfo {volume} {497}},\ \bibinfo {pages}
  {598} (\bibinfo {year} {2013})}\BibitemShut {NoStop}%
\bibitem [{\citenamefont {Kim}\ \emph {et~al.}(2018)\citenamefont {Kim},
  \citenamefont {Leconte}, \citenamefont {Chittari}, \citenamefont {Watanabe},
  \citenamefont {Taniguchi}, \citenamefont {MacDonald}, \citenamefont {Jung},\
  and\ \citenamefont {Jung}}]{Kim2018a}%
  \BibitemOpen
  \bibfield  {author} {\bibinfo {author} {\bibfnamefont {H.}~\bibnamefont
  {Kim}}, \bibinfo {author} {\bibfnamefont {N.}~\bibnamefont {Leconte}},
  \bibinfo {author} {\bibfnamefont {B.~L.}\ \bibnamefont {Chittari}}, \bibinfo
  {author} {\bibfnamefont {K.}~\bibnamefont {Watanabe}}, \bibinfo {author}
  {\bibfnamefont {T.}~\bibnamefont {Taniguchi}}, \bibinfo {author}
  {\bibfnamefont {A.~H.}\ \bibnamefont {MacDonald}}, \bibinfo {author}
  {\bibfnamefont {J.}~\bibnamefont {Jung}}, \ and\ \bibinfo {author}
  {\bibfnamefont {S.}~\bibnamefont {Jung}},\ }\href {\doibase
  10.1021/acs.nanolett.8b03423} {\bibfield  {journal} {\bibinfo  {journal}
  {Nano Lett.}\ }\textbf {\bibinfo {volume} {18}},\ \bibinfo {pages} {7732}
  (\bibinfo {year} {2018})}\BibitemShut {NoStop}%
\bibitem [{\citenamefont {Jones}\ \emph {et~al.}(2017)\citenamefont {Jones},
  \citenamefont {Bahamon}, \citenamefont {{Castro Neto}},\ and\ \citenamefont
  {Pereira}}]{Jones:2017}%
  \BibitemOpen
  \bibfield  {author} {\bibinfo {author} {\bibfnamefont {G.~W.}\ \bibnamefont
  {Jones}}, \bibinfo {author} {\bibfnamefont {D.~A.}\ \bibnamefont {Bahamon}},
  \bibinfo {author} {\bibfnamefont {A.~H.}\ \bibnamefont {{Castro Neto}}}, \
  and\ \bibinfo {author} {\bibfnamefont {V.~M.}\ \bibnamefont {Pereira}},\
  }\href {\doibase 10.1021/acs.nanolett.7b01663} {\bibfield  {journal}
  {\bibinfo  {journal} {Nano Lett.}\ }\textbf {\bibinfo {volume} {17}},\
  \bibinfo {pages} {5304} (\bibinfo {year} {2017})}\BibitemShut {NoStop}%
\bibitem [{\citenamefont {M{\"{u}}ller}(1992)}]{Muller1992}%
  \BibitemOpen
  \bibfield  {author} {\bibinfo {author} {\bibfnamefont {J.~E.}\ \bibnamefont
  {M{\"{u}}ller}},\ }\href {\doibase 10.1103/PhysRevLett.68.385} {\bibfield
  {journal} {\bibinfo  {journal} {Phys. Rev. Lett.}\ }\textbf {\bibinfo
  {volume} {68}},\ \bibinfo {pages} {385} (\bibinfo {year} {1992})}\BibitemShut
  {NoStop}%
\bibitem [{\citenamefont {Chou}\ \emph {et~al.}(2019)\citenamefont {Chou},
  \citenamefont {Lin}, \citenamefont {{Das Sarma}},\ and\ \citenamefont
  {Nandkishore}}]{Chou2019}%
  \BibitemOpen
  \bibfield  {author} {\bibinfo {author} {\bibfnamefont {Y.-Z.}\ \bibnamefont
  {Chou}}, \bibinfo {author} {\bibfnamefont {Y.-P.}\ \bibnamefont {Lin}},
  \bibinfo {author} {\bibfnamefont {S.}~\bibnamefont {{Das Sarma}}}, \ and\
  \bibinfo {author} {\bibfnamefont {R.~M.}\ \bibnamefont {Nandkishore}},\
  }\href {\doibase 10.1103/PhysRevB.100.115128} {\bibfield  {journal} {\bibinfo
   {journal} {Phys. Rev. B}\ }\textbf {\bibinfo {volume} {100}},\ \bibinfo
  {pages} {115128} (\bibinfo {year} {2019})}\BibitemShut {NoStop}%
\end{thebibliography}%

% ------------------------------------------------------------------------------
% APPENDIX BEGINS
% ------------------------------------------------------------------------------
\appendix
\begin{widetext}
 
\section{Bosonization conventions}

According to the bosonization approach \cite{Giamarchi2003,Fradkin2013},
the fermion field from the $l$-th wire of array $j$ with spin $\sigma$ and 
direction $\mu$ ($+1$ for right and $-1$ for left) reads
\begin{equation}
\psi^{j}_{l,\mu,\sigma} \propto \frac{1}{\sqrt{2\pi \alpha}}
e^{i \mu 2\sqrt{\pi} \phi^j_{l,\mu,\sigma}},
\end{equation}
where $\alpha$ is the length cut-off. The 
bosonic fields are related by:
\begin{subequations}\label{eq:bosonic-fields}
\begin{align}
& \phi^j_{l,\sigma} = \phi^j_{l,+,\sigma} + \phi^j_{l,-,\sigma}, \\
& \theta^j_{l,\sigma} = -\phi^j_{l,+,\sigma} + \phi^j_{l,-,\sigma}, \\
& \phi^j_{l,c/s} = \frac{1}{\sqrt{2}} ( \phi^j_{l,\uparrow} \pm 
\phi^j_{l,\downarrow} ), \\
& \theta^j_{l,c/s} = \frac{1}{\sqrt{2}} ( \theta^j_{l,\uparrow} \pm 
\theta^j_{l,\downarrow} ).
\end{align}
\end{subequations}
The subscript $c/s$ in \Eqref{eq:bosonic-fields}(c,d) stands for the 
charge/spin sector. The bosonic fields are related to the electron densities 
through:
\begin{subequations}\label{eq:density}
\begin{align}
& \frac{1}{\sqrt{\pi}} \partial_{x_j} \phi^j_{l,\mu,\sigma} = 
\rho^j_{l,\mu,\sigma} , \\
& \frac{1}{\sqrt{\pi}} \partial_{x_j} \phi^j_{l,c/s} = \frac{1}{\sqrt{2}}
(\rho^j_{l,+,c/s} + \rho^j_{l,-,c/s} ) , \\
& \frac{1}{\sqrt{\pi}} \partial_{x_j} \theta^j_{l,c/s} = \frac{1}{\sqrt{2}}
( -\rho^j_{l,+,c/s} + \rho^j_{l,-,c/s} ) ,
\end{align}
\end{subequations}
where $x_j$ is the variable along the direction of $j$-th array and $\rho_{c/s}$ 
stands for the charge/spin density, i.e., $\rho_{c/s} = \rho_{\uparrow} \pm 
\rho_{\downarrow}$.

\section{Bosonized form of the couplings}

Once we know the bosonized form of the fermion fields, it is straightforward to 
write all the couplings in terms of the bosonic fiels defined in the previous 
section. For the intra-array couplings, we consider an array of parallel 
wires along the $x$ direction [see the schematic of array-$1$ in 
\Fref{fig:schematic}(a) as an example]: The Hamiltonian terms 
describing the coupling between wires separated by a distance $n d$ are:
\begin{subequations}\label{eq:intra-coupling}
\begin{align}
H^j_{c,n} &= \mathcal{V}_n \sum_{l} \sum_{\mu,\sigma,\sigma'} \int dx
\psi^{j,\dagger}_{l,\mu,\sigma}\psi^j_{l,-\mu,\sigma}
\psi^{j,\dagger}_{l+n,-\mu,\sigma'}\psi^j_{l+n,\mu,\sigma'} \nonumber \\
&= \frac{2 \mathcal{V}_n}{(\pi \alpha)^2} \sum_{l} \int dx \cos\left[
\sqrt{2\pi}(\phi^j_{l,c}-\phi^j_{l+n,c}) \right]
\cos(\sqrt{2\pi}\phi^j_{l,s}) \cos(\sqrt{2\pi}\phi^j_{l+n,s}) \\[1.5em]
H^j_{sc,n} &= \mathcal{J}_n \sum_{l,\mu,\nu} \int dx 
\psi^{j,\dagger}_{l,\mu,\uparrow}
\psi^{j,\dagger}_{l,-\mu,\downarrow} \psi^j_{l+n,\nu,\downarrow} 
\psi^j_{l+n,-\nu,\uparrow} + \hc
\nonumber \\
&= \frac{2 \mathcal{J}_n}{(\pi \alpha)^2} \sum_l \int dx
\cos\left[ \sqrt{2\pi}(\theta^j_{l,c}-\theta^j_{l+n,c}) \right]
\cos(\sqrt{2\pi}\phi^j_{l,s}) \cos(\sqrt{2\pi}\phi^j_{l+n,s}) \\[1.5em]
H^j_{h} &= t_{\perp} \sum_{l} \sum_{\mu,\sigma} \int dx
\psi^{j,\dagger}_{l,\mu,\sigma}\psi^j_{l+1,\mu,\sigma} + \hc \nonumber \\
&= \frac{2 t_{\perp}}{\pi \alpha} \sum_{l} \int dx
\cos\left[ \sqrt{\frac{\pi}{2}} 
(\phi^j_{l,c}-\theta^j_{l,c}-\phi^j_{l+1,c}+\theta^j_{l+1,c} ) \right]
\cos\left[ \sqrt{\frac{\pi}{2}} 
(\phi^j_{l,s}-\theta^j_{l,s}-\phi^j_{l+1,s}+\theta^j_{l+1,s} ) \right] \nonumber 
\\
&\ \times
\cos\left[ \sqrt{\frac{\pi}{2}} 
(\phi^j_{l,c}+\theta^j_{l,c}-\phi^j_{l+1,c}-\theta^j_{l+1,c} ) \right]
\cos\left[ \sqrt{\frac{\pi}{2}} 
(\phi^j_{l,s}+\theta^j_{l,s}-\phi^j_{l+1,s}-\theta^j_{l+1,s} ) \right]
\end{align}
\end{subequations}
Meanwhile, the inter-array couplings at a crossing point (e.g., the 
intersection of the $l$-th wire from array $k$ and the $m$-th from array $j$) 
can be written in the form:
\begin{subequations}\label{eq:inter-coupling}
\begin{align}
H^{k,j}_{cdw} &= \mathcal{V}_0 \sum_{\mu,\nu,\sigma,\sigma'}
\psi^{k \dagger}_{l,\mu,\sigma}\psi^k_{l,-\mu,\sigma}
\psi^{j \dagger}_{m,\nu,\sigma'}\psi^j_{m,-\nu,\sigma'} \nonumber \\
&= \frac{4 \mathcal{V}_0}{(\pi \alpha)^2} \cos(\sqrt{2\pi} \phi^k_{l,c}) 
\cos(\sqrt{2\pi} \phi^k_{l,s})
\cos(\sqrt{2\pi} \phi^j_{m,c}) \cos(\sqrt{2\pi} \phi^j_{m,s}) \\[1.5em]
H^{k,j}_{sc} &= \mathcal{J}_0 \sum_{\mu,\nu}
\psi^{k \dagger}_{l,\mu,\uparrow}\psi^{k \dagger}_{l,-\mu,\downarrow}
\psi^{j}_{m,\nu,\downarrow}\psi^j_{m,-\nu,\uparrow} + \hc \nonumber \\
&= \frac{4 \mathcal{J}_0}{(\pi \alpha)^2} e^{i \sqrt{2\pi} ( \theta^k_{l,c} - 
\theta^j_{m,c} )}
\cos(\sqrt{2\pi} \phi^l_{l,s}) 
\cos(\sqrt{2\pi} \phi^j_{m,s}) + \hc \\[1.5em]
H^{k,j}_{h} &= t \sum_{\mu,\nu,\sigma}
\psi^{k \dagger}_{l,\mu,\sigma} \psi^{j}_{m,\nu,\sigma}
+ \hc \nonumber \\
&= \frac{t}{\pi \alpha} \sum_{\sigma}
e^{ i \sqrt{\frac{\pi}{2}} (\theta^k_{l,c} + \sigma \theta^k_{l,s}
-\theta^j_{m,c} - \sigma \theta^j_{m,s}) }
\cos\left[ \sqrt{\frac{\pi}{2}} (\phi^k_{l,c} + \sigma \phi^k_{l,s}) \right]
\cos\left[ \sqrt{\frac{\pi}{2}} (\phi^j_{m,c} + \sigma \phi^j_{m,s}) \right]
 + \hc 
\end{align}
\end{subequations}
Having obtained the identities above, one can proceed with the perturbative RG 
calculation to explore the potential instabilities of the system.

\end{widetext}

\end{document}